\documentclass[manuscript]{acmart}

\usepackage{tikz}
\usetikzlibrary{arrows.meta, positioning, calc, fit}
\usepackage{empheq}
\usepackage{float}
\usepackage{adjustbox} 
\usepackage{amsthm}
\newtheorem{globaltheorem}{Global Theorem}
\usepackage{subcaption}
\usepackage{enumitem}
\usepackage{multirow}
\usepackage{colortbl}
\usepackage{listings}

\usepackage{acronym}

\usepackage[ruled,vlined]{algorithm2e}
\SetKwBlock{Procedure}{Procedure}{End Procedure}

\acrodef{QAP}{Quadratic Arithmetic Program}
\acrodef{R1CS}{Rank 1 Constraint Systems}
\acrodef{MPC}{Multi-Party Computation}
\acrodef{TC}{Tornado Cash}
\acrodef{DLTs}{Distributed Ledger Technologies}
\acrodef{HSDLs}{High-Scalability Distributed Ledgers}
\acrodef{ZKPs}{Zero-Knowledge Proof}
\acrodef{FS}{Front-end Server}
\acrodef{SNARKs}{Succinct Non-Interactive Argument of Knowledge}
\acrodef{STARKs}{Scalable Transparent Argument of Knowledge}
\acrodef{XRPL}{XRP Ledger}
\acrodef{TPS}{Transactions Per Second}
\acrodef{BFT}{Byzantine Fault Tolerance}
\acrodef{PBFT}{Practical Byzantine Fault Tolerance}
\acrodef{PoW}{Proof-of-Work}
\acrodef{PoET}{Proof-of-Elapsed-Time}
\acrodef{PoSpace}{Proof-of-Space}
\acrodef{PoS}{Proof-of-Stake}
\acrodef{PoA}{Proof-of-Authority}
\acrodef{RCC}{Resilient Concurrent Consensus}
\acrodef{DAG}{Directed Acyclic Graph}
\acrodef{TEE}{Trusted Execution Environment}
\acrodef{RPCA}{Ripple Protocol Consensus Algorithm}
\acrodef{UNL}{Unique Node List}
\acrodef{TVL}{Total Value Locked}
\acrodef{QSP}{Quadratic Span Program}
\acrodef{CRS}{Common Reference String}
\acrodef{FFT}{Fast Fourier Transform}
\acrodef{API}{Application Programming Interface}
\acrodef{EVM}{\emph{Ethereum} Virtual Machine}
\acrodef{SDK}{Software Development Kit}
\acrodef{EdDSA}{Edwards-curve Digital Signature Algorithm}
\acrodef{ECDSA}{Elliptic Curve Digital Signature Algorithm}
\acrodef{WASM}{WebAssembly}
\acrodef{RAM}{Random Access Memory}
\acrodef{CPU}{Central Processing Unit}
\acrodef{L2}{Layer 2}
\acrodef{L1}{Layer 1}
\acrodef{AWS}{Amazon Web Services}
\acrodef{OFAC}{Office of Foreign Assets Control}
\acrodef{Dapps}{Decentralized Applications}
\acrodef{FCFS}{First Come First Serve}
\acrodef{DoS}{Denial of Service}

\usepackage{siunitx}
\usepackage{xcolor}
\usepackage{csquotes}
\usepackage{hyperref}
\lstdefinelanguage{JavaScript}{
  keywords={typeof, new, true, false, catch, function, return, null, catch, const, for, let, 
    switch, var, if, in, while, do, else, case, break},
  keywordstyle=\color{purple},
  ndkeywords={class, export, boolean, throw, implements, import, this},
  ndkeywordstyle=\color{yellow}\bfseries,
  identifierstyle=\color{blue},
  sensitive=false,
  comment=[l]{//},
  morecomment=[s]{/*}{*/},
  commentstyle=\color{gray}\ttfamily,
  stringstyle=\color{red}\ttfamily,
  morestring=[b]',
  morestring=[b]"
}

\lstdefinelanguage{Circom}{
  keywords={signal, input, output, template, component, for, var},
  keywordstyle=\color{purple},
  ndkeywords={class, export, boolean, throw, implements, import, this},
  ndkeywordstyle=\color{yellow}\bfseries,
  identifierstyle=\color{blue},
  sensitive=false,
  comment=[l]{//},
  morecomment=[s]{/*}{*/},
  commentstyle=\color{gray}\ttfamily,
  stringstyle=\color{red}\ttfamily,
  morestring=[b]',
  morestring=[b]"
}

\lstdefinelanguage{Solidity}{
  keywords={pragma, import, contract, function, return, event, emit, mapping, address, uint, string, bool, require, view, payable, external, internal, private, public, pure, constant, memory, storage, enum, struct, constructor, modifier, interface, override, library, abstract, event},
  keywordstyle=\color{purple},
  ndkeywords={true, false},
  ndkeywordstyle=\color{yellow}\bfseries,
  identifierstyle=\color{blue},
  sensitive=true,
  comment=[l]{//},
  morecomment=[s]{/*}{*/},
  commentstyle=\color{gray}\ttfamily,
  stringstyle=\color{red}\ttfamily,
  morestring=[b]',
  morestring=[b]"
}

\AtBeginDocument{%
  }

\begin{document}

\title{Benchmarking ZK-Friendly Hash Functions and SNARK Proving Systems for EVM-compatible Blockchains}

\author{Hanze Guo}
\affiliation{%
  \institution{The DLT Science Foundation}
  \country{UK}}
\email{guohanze@connect.hku.hk}

\author{Yebo Feng}
\affiliation{%
  \institution{Nanyang Technological University}
  \country{Singapore}}
\email{yebo.feng@ntu.edu.sg}

\author{Cong Wu}
\affiliation{%
  \institution{The University of Hong Kong}
  \country{China}}
\email{congwu@hku.hk}

\author{Zengpeng Li}
\affiliation{%
  \institution{Shandong University}
  \country{China}}
\email{zengpeng@email.sdu.edu.cn}

\author{Jiahua Xu}
\affiliation{%
  \institution{University College London, The DLT Science Foundation}
  \country{UK}}
\email{jiahua.xu@ucl.ac.uk}

\begin{abstract}
With the rapid development of \ac{ZKPs}, particularly \ac{SNARKs}, benchmarking various ZK development tools has proven to be a valuable and necessary task. ZK-friendly hash functions, as foundational algorithms widely used in blockchain applications, have garnered significant attention. Therefore, conducting comprehensive benchmarking and comparative evaluations of the implementations of these continually evolving algorithms in ZK circuits is a promising and challenging endeavor. Additionally, we focus on one of the popular ZKP applications—privacy-preserving transaction protocols—aiming to leverage \ac{SNARKs}' cost-efficiency potential combined with the idea of \enquote{batch processing} to overcome high on-chain costs and compliance challenges these applications commonly face.

To this end, we benchmarked three SNARK proving systems and five ZK-friendly hash functions, including our self-developed circuit templates for \emph{Poseidon2}, \emph{Neptune}, and \emph{GMiMC}, on \emph{bn254} curve within \emph{circom-snarkjs} framework. Additionally, we introduced the role of \enquote{sequencer} in our SNARK-based privacy-preserving transaction scheme to achieve efficiency improvement and flexible auditing. We conducted privacy and security analyses, as well as implementation and evaluation on \ac{EVM}-compatible chains. The benchmarking results indicate that \emph{Poseidon} and \emph{Poseidon2} exhibit superior memory consumption and runtime during the proof generation phase under \emph{Groth16}. Moreover, compared to the baseline, the implementation with \emph{Poseidon2} not only results in faster proof generation but also reduces on-chain costs by 73\% on EVM chains and nearly 26\% on \emph{Hedera}. Our work provides a benchmark for ZK-friendly hash functions and popular ZK development tools, while also offering exploratory research on cost efficiency and compliance in ZKP-based privacy-preserving transaction protocols.

\end{abstract}


\maketitle

\section{Introduction} \label{intro}
Blockchain technology has transformed value transfer systems by enabling decentralized, transparent, and secure transactions without third-party intermediaries. As blockchain adoption grows, \ac{ZKPs} have become essential for enhancing blockchain performance~\cite{scroll, Starknet}. With rapid advancements in ZKPs, particularly \ac{SNARKs}, benchmarking various ZK development tools, especially at the ZK circuit level, has become crucial~\cite{zkbench, benchmarkingcircom}. ZK-friendly hash functions are foundational for cryptographic constructs like Merkle trees, impacting the efficiency of higher-level applications. Therefore, we first focus on implementing these widely-followed algorithms in ZK circuits, aiming to provide ZK developers with a comprehensive and detailed evaluation and benchmark.

We explore the potential of ZKPs in blockchain privacy-preserving transaction protocols~\cite{Meiklejohn2018, Tairi2021, RuffingCoinshuffle}. Inspired by \emph{Zerocash}~\cite{zerocash}, these protocols obfuscate transaction chains using multiple mixer pools with fixed transaction amounts. The privacy strength, determined by the anonymity set size, depends on the number of transactions in the protocol~\cite{wang2023on}. A common approach to expanding the anonymity set is framing incentive mechanisms to attract more users~\cite{LeAMR, TornadoCash, miximus}. However, research by Wang et al.~\cite{wang2023on} shows that these mechanisms often attract users who are more interested in rewards than privacy, which presents certain limitations. Additionally, private transactions are costly; for example, a single fixed-amount transaction can consume around 1.2 million gas, equivalent to approximately 39 USD, which is nearly five times the average transaction fee of 8 USD.\footnote{At the time of writing, 1 ETH = 3,251 USD, and the gas price on the \emph{Ethereum} mainnet is 10 Gwei.} This high cost raises the participation threshold for privacy-seeking users. Our scheme aims to lower this threshold, expanding the anonymity set by moving resource-intensive transaction processing off-chain through a \enquote{sequencer}, allowing verifiable computation on-chain without revealing inputs or intermediate stages~\cite{xiangan2024}.

The introduction of the \enquote{sequencer} role addresses uncontrolled censorship resistance in ZKP-based privacy protection schemes. \emph{\ac{TC}} was sanctioned by the U.S. Department of the Treasury's \ac{OFAC} on August 8, 2022, for allegedly facilitating money laundering~\cite{wang2023on}, causing its project team to cease operations. However, because \emph{TC} is built on a decentralized smart contract platform, users could still use \emph{TC}'s services despite regulatory actions. A report indicated that as of March 2024, over 283 million USD in crypto assets had flowed into \emph{TC}'s mixer pools~\cite{wired2024}, suggesting centralized regulation fails to prevent illicit use of such \ac{Dapps}. Our scheme introduces a trusted \enquote{sequencer} to implement necessary censorship measures by pre-screening interacting addresses, similar to the blacklist feature in \emph{Haze}~\cite{2023haze}, thereby preventing unauthorized transactions and managing illicit funds.

The paper outlines cryptographic preliminaries in \autoref{bg}. Section \ref{bench} evaluates the performance of three SNARKs protocols and five popular ZK-friendly hash algorithms: \emph{MiMC}~\cite{albrecht2016mimc}, \emph{GMiMC}~\cite{ben2020}, \emph{Poseidon}~\cite{grassi2021p1}, \emph{Poseidon2}~\cite{grass2023}, and \emph{Neptune}~\cite{alan2023}. Section \ref{exploration} discusses the overflow, threat model, methods for batch processing approach, the introduced compliance and privacy evaluation. Section \ref{impl} describes the implementation process, including \emph{circom}-written circuits~\cite{repocircom} and \emph{solidity}-written smart contracts. We replicate \emph{TC}'s performance as a baseline to evaluate optimized prover time and cost in \emph{Ethereum}, \emph{Hedera}, and \emph{BNB Chain}. Our contributions are summarized as follows:

\begin{itemize}
    \item We propose a batch processing and SNARKs-based solution to address high on-chain costs and uncontrolled compliance in privacy-preserving blockchain transactions. Using advanced cryptographic hash functions like \emph{Poseidon2}~\cite{grass2023}, we significantly reduce on-chain costs by over 70\% without sacrificing longer proof generation times compared to baseline protocols.
    
    \item We develop more advanced and extensive instances of hash functions for \emph{circom} and provided comprehensive benchmarks of five ZK-friendly hash functions using three proving systems within \emph{snarkjs} on \emph{bn254}. This benchmark covers the setup, prove, and verify stages, as well as metrics such as runtime, and \ac{RAM} consumption.
\end{itemize}

\section{Background} \label{bg}

\subsection{Notations} \label{subsubsec: notations}
We use $x$ represents a single variable, while bold letters $\mathbf{x}$ represent vectors composed of multiple variables. $\mathbf{x}_i$ denotes the $i$-th element in $\mathbf{x}$. We use \( \mathbf{x}[i:j] \) to denote the slice of \( \mathbf{x} \), representing \( \mathbf{x}_i, \mathbf{x}_{i+1}, \ldots, \mathbf{x}_j \). Let $\mathbb{Z}_p$ denote a scalar finite field over a prime order $\mathbb{F}_p$ with a base field \(\mathbb{F}_q\). $H_{\it Pedersen}: \mathbb{Z}_p^2 \rightarrow \mathbb{Z}_p$ represents the \emph{Pedersen} hash function defined in~\cite{pedersenhash} for an input length of 2. We denote the \emph{MiMC} hash function as $H_{\it MiMC}: \mathbb{Z}_p^* \rightarrow \mathbb{Z}_p$, as defined in~\cite{albrecht2016mimc} for any valid input length, and the \emph{Poseidon2} hash function as $H_{\it Poseidon2}: \mathbb{Z}_p^* \rightarrow \mathbb{Z}_p$, as defined in~\cite{grass2023} for any valid input length.

\subsection{Merkle Tree} \label{subsec: merkletree}
A Merkle tree~\cite{merkle1987digital} is a data structure used for vector commitment schemes, often employed in blockchain. Variants include the incremental Merkle tree for efficient user-submitted commitment verification in \emph{TC}~\cite{TornadoCash}. We use this variant in \autoref{exploration}, characterized by initializing leaf nodes to zero\footnote{Typically, the initial value can be any constant.} and having a fixed depth $d_{\it MT}$. During state transitions, the leaf nodes are updated in a fixed left-to-right order, sequentially replacing zero values with new commitments. These features result in a more efficient algorithm compared to others. \autoref{equation1} shows the generation of a binary \(\mathcal{M}\)\footnote{Unless specified, the Merkle trees mentioned in this paper are assumed to be binary, where each node's value is derived from its left and right children using a specific hash function.} using a hash function \(\mathcal{H}\) with an ordered array of leaf nodes \(\mathbf{L}\):
\begin{equation} \label{equation1}
    (\mathbf{L}, \mathcal{H}) \rightarrow \mathcal{M}, \mathcal{M}\ni \left( \mathbf{L}, r, \pi \right),
\end{equation}
where some key attributes of \(\mathcal{M}\) are listed, including the root node's hash \(r\), the leaf node vector \(\mathbf{L}\), and the path set \(\pi\) corresponding to \(\mathbf{L}\), where \(\pi_i\) corresponds to \(\mathbf{L}_i\). Since \(\mathcal{H}\) is sequence-sensitive, \(\pi_i\) includes not only a set of necessary hash paths for computing \(r\) but also their positional relationships. This algorithm has \(O(n)\) time and space complexity for a full binary Merkle tree relative to the size \(n\) of \(\mathbf{L}\). For incremental Merkle trees, the time and space complexity are both reduced to \(O(\log_2 n)\). \autoref{equation2} shows the state transition process of a sparse \(\mathcal{M}\) generated by \autoref{equation1}, replacing the original leaf node \(\mathbf{L}_i\) with a new commitment \(\mathcal{C}\):
\begin{equation} \label{equation2}
    (\mathcal{C}, \pi_i) \rightarrow \mathcal{M}^*, \mathcal{M}^* \ni (\mathbf{L}^*, r^*, \pi^*),
\end{equation}
updating with \(O(\log_2 n)\) time complexity. When $\mathcal{M}$ is an incremental Merkle tree, the space complexity reduces from \(O(n)\) to \(O(\log_2 n)\). \autoref{equation3} represents verifying the existence of the new commitment $\mathcal{C}$ in $\mathcal{M}^*$:
\begin{equation} \label{equation3}
    \it (\mathcal{C}, r^*, \pi^*_i) \in \mathcal{M}^* \rightarrow \{\it True, \it False\},
\end{equation}
verifying with \(O(\log_2 n)\) time complexity. The space complexity reduces from \(O(n)\) to \(O(\log_2 n)\) when $\mathcal{M}$ is an incremental Merkle tree.

\subsection{Zero-Knowledge Proof} \label{subsec: zkp}
In ZKP protocols, a \emph{prover} aims to convince a \emph{verifier} that a certain statement is correct through an instance \( \mathbf{x} \), and the evidence or secret information \emph{prover} holds to prove the statement is called the witness \( \mathbf{w} \). \( \mathbf{w} \) and \( \mathbf{x} \) are in an NP relation \( \mathcal{R} \), i.e., \( (\mathbf{x},\mathbf{w}) \in \mathcal{R} \), which can be verified by \emph{verifier}, where \( \mathcal{R} \) is known to both \emph{verifier} and \emph{prover}~\cite{zkbench}. This process possesses the properties of \emph{completeness}, \emph{soundness}, and \emph{zero-knowledge}, which we introduce in the three stages of ZKP described below.

\begin{enumerate}
    \item \(\it Setup(\lambda) \rightarrow \mathbf{pp} \): Generate a set of public parameters \( \mathbf{pp} \) that include the keys used for proving and verifying, with a security parameter \( \lambda \).
    \item \(\it Prove(\mathbf{x}, \mathbf{w}, \mathbf{pp}) \rightarrow \pi \): \emph{prover} generates the ZKP \(\pi\). This process reflects \emph{soundness}, meaning \emph{prover} cannot produce \( \pi \) by any cheating means if they only know the public \( \mathbf{x} \) and \( \mathbf{pp} \) without knowing the secret \( \mathbf{w} \).
    \item \(\it Verify(\mathbf{x}, \pi, \mathbf{pp}) \rightarrow \{\it True, \it False\} \): \emph{verifier} verifies \(\pi\). This process demonstrates \emph{completeness} and \emph{zero-knowledge}, meaning \( \it Verify(\mathbf{x},\pi,\mathbf{pp}) \rightarrow \it True \) convinces \emph{verifier}, and \emph{verifier} only learns \( \{\it True, \it False\} \) without gaining any other information.
\end{enumerate}

Our work focuses on pre-processing SNARK protocols, which utilize a setup algorithm to pre-process the relation \( \mathcal{R} \), reducing the computational burden on verifiers for different \( \mathbf{x} \) scenarios. We detail the three proof systems, \emph{Groth16}~\cite{Groth16}, \emph{Plonk}~\cite{gabizon2019Plonk}, and \emph{Fflonk}~\cite{gabizon2021Fflonk}, supported in \emph{snarkjs} below, specifically highlighting their differences in universality of setup and arithmetization.

\paragraph{Universality of Setup}
\emph{Groth16}~\cite{Groth16} has the advantage of efficiently generating proofs, making it very suitable for on-chain verification scenarios. However, \emph{Groth16} requires a trusted setup for each circuit. \emph{Plonk}~\cite{gabizon2019Plonk} was specifically designed to address this limitation. \emph{Plonk} only requires a single universal trusted setup for circuits of any size within the range, eliminating the need for re-setup. \emph{Fflonk}~\cite{gabizon2021Fflonk} is variant of \emph{Plonk}. This variant generally increases the computational resource costs for proof generation~\cite{ambrona22}. According to Jordi~\cite{jordi23}, the proof generation time for \emph{Fflonk} is approximately ten times slower than \emph{Groth16}. However, \emph{Fflonk} is still suitable for specific scenarios, such as reducing on-chain verification costs, where it can be about a little cheaper than \emph{Groth16} and about 30\% cheaper than \emph{Plonk} in terms of gas costs.

\paragraph{R1CS vs Plonkish Arithmetization} We outline the specific methods of arithmetization, which is the method of converting arithmetic circuits into polynomial constraints. \emph{Groth16} algorithm is based on \ac{QAP} problem. All computational circuits can be converted into \ac{QAP} problems via \ac{R1CS}. \emph{Plonk}~\cite{gabizon2019Plonk}, as mentioned above, avoids the need for a one-time trusted setup by designing subsequent constraint systems and problem compression methods. Plonkish arithmetization is a unique arithmetization method of \emph{Plonk} proving system. Before Plonkish, the widely adopted circuit representation form was \ac{R1CS}~\cite{pinocchio16, bulletproofs18}.

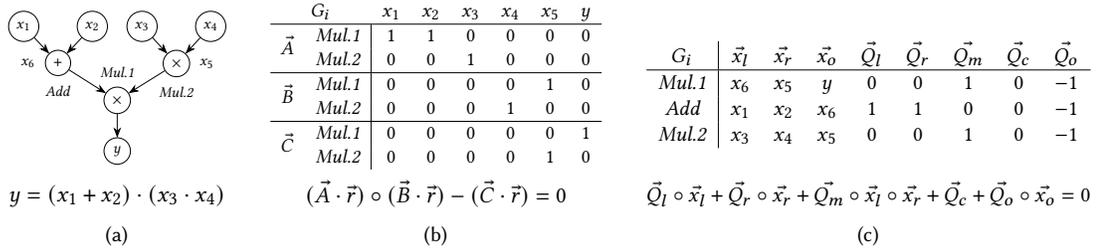
\begin{figure}[h]
    \centering
        \begin{subfigure}[b]{0.2\textwidth}
        \centering
        \resizebox{\textwidth}{!}{
        \begin{tikzpicture}[node distance=0.2cm, auto, ->, >=Stealth, scale=0.8, every node/.append style={transform shape}]
        \node (x1) [draw, circle] {$x_1$};
        \node (x2) [right=0.8cm of x1, draw, circle] {$x_2$};
        \node (x3) [right=0.4cm of x2, draw, circle] {$x_3$};
        \node (x4) [right=0.8cm of x3, draw, circle] {$x_4$};
            
        \node (add) [below=of $(x1)!0.5!(x2)$, yshift=-0.3cm, draw, circle] {$+$};
        \node (mul1) [below=of $(x3)!0.5!(x4)$, yshift=-0.3cm, draw, circle] {$\times$};
        \node (mul2) [below=of $(add)!0.5!(mul1)$, yshift=-0.3cm, draw, circle] {$\times$};
        \node (y) [below=of mul2, yshift=-0.3cm, draw, circle] {$y$};

        \draw[->] (x1) -- (add);
        \draw[->] (x2) -- (add);
        \draw[->] (add) -- (mul2);
        \draw[->] (x3) -- (mul1);
        \draw[->] (x4) -- (mul1);
        \draw[->] (mul1) -- (mul2);
        \draw[->] (mul2) -- (y);
            
        \node[below=0.1cm of mul1] {\emph{Mul.2}};
        \node[above=0.1cm of mul2] {\emph{Mul.1}};
        \node[below=0.1cm of add] {\emph{Add}};
        \node[right=0.1cm of mul1] {$x_5$};
        \node[left=0.1cm of add] {$x_6$};
        \end{tikzpicture}
        }
        \par\vspace{0.1cm}
        \resizebox{\textwidth}{!}{
        $y=(x_1 + x_2) \cdot (x_3 \cdot x_4)$
        }
        \caption{}
        \label{dag}
    \end{subfigure}
    \hspace{0.02\textwidth}
    \begin{subfigure}[b]{0.3\textwidth}
        \centering
        \resizebox{\textwidth}{!}{
        $\begin{array}{cc|cccccc}
        \multicolumn{2}{c}{G_i} & x_1 & x_2 & x_3 & x_4 & x_5 & y \\ \hline
        \multirow{2}{*}{$\vec{A}$} & \emph{Mul.1} & 1 & 1 & 0 & 0 & 0 & 0 \\
                           & \emph{Mul.2} & 0 & 0 & 1 & 0 & 0 & 0 \\ \hline
        \multirow{2}{*}{$\vec{B}$} & \emph{Mul.1} & 0 & 0 & 0 & 0 & 1 & 0 \\
                           & \emph{Mul.2} & 0 & 0 & 0 & 1 & 0 & 0 \\ \hline
        \multirow{2}{*}{$\vec{C}$} & \emph{Mul.1} & 0 & 0 & 0 & 0 & 0 & 1 \\
                           & \emph{Mul.2} & 0 & 0 & 0 & 0 & 1 & 0 \\
        \end{array}$
        }
        %
        \par\vspace{0.1cm}
        \resizebox{0.8\textwidth}{!}{
        $(\vec{A} \cdot \vec{r}) \circ (\vec{B} \cdot \vec{r}) - (\vec{C} \cdot \vec{r}) = 0$
        }
        \caption{}
        \label{r1cs}
    \end{subfigure}
    \hspace{0.02\textwidth}
    \begin{subfigure}[b]{0.4\textwidth} 
        \centering
        \resizebox{\textwidth}{!}{
        $\begin{array}{c|cccccccc}
           G_i & \vec{x_l} & \vec{x_r} & \vec{x_o} & \vec{Q_l} & \vec{Q_r} & \vec{Q_m} & \vec{Q_c} & \vec{Q_o} \\ \hline
           \emph{Mul.1} & x_6 & x_5 & y & 0 & 0 & 1 & 0 & -1 \\
           \emph{Add} & x_1 & x_2 & x_6 & 1 & 1 & 0 & 0 & -1 \\
           \emph{Mul.2} & x_3 & x_4 & x_5 & 0 & 0 & 1 & 0 & -1 \\
        \end{array}$
        }
        \par\vspace{0.4cm}
        \resizebox{\textwidth}{!}{
        $\vec{Q_l} \circ \vec{x_l} + \vec{Q_r} \circ \vec{x_r} + \vec{Q_m} \circ \vec{x_l} \circ \vec{x_r} + \vec{Q_c} + \vec{Q_o} \circ \vec{x_o} = 0$
        }
        \caption{}
        \label{plonkish}
    \end{subfigure}
    \caption{The arithmetic circuit $C(x_1, x_2, x_3, x_4) = (x_1 + x_2) \cdot (x_3 \cdot x_4)$ over $\mathbb{F}_q$ in different representations. Fig. \ref{dag} shows Graph representation. Fig. \ref{r1cs} shows \ac{R1CS} representation. Fig. \ref{plonkish} shows Plonkish representation.}
    \label{arithmetization}
\end{figure}

Arithmetic circuits as the frontend for SNARKs can be represented as a directed acyclic graph~\cite{zkbench}. \autoref{arithmetization} illustrates three representations of a circuit containing two multiplication gates and one addition gate, namely the graph, R1CS, and Plonkish. This example demonstrates two arithmetic differences. In R1CS, the focus is on multiplication gates, using three matrices $\vec{A}$, $\vec{B}$, and $\vec{C}$ to define which variables are connected to the left input, right input, and output of each multiplication gate in the circuit. As shown in \autoref{arithmetization}, the addition gate in the circuit does not increase the number of rows in the three matrices, making the encoding method of R1CS circuits have minimal performance impact on the prover due to addition operations, with a relatively simple number of constraints. On the other hand, \emph{Plonk}'s encoding scheme requires encoding both addition and multiplication gates, which seems to reduce prover performance due to the increased number of constraints. This is demonstrated in detail in \autoref{subsec: circuit power}. However, Plonkish introduces gates beyond multiplication and addition, including XOR gates and even custom gates that support any polynomial relationship between inputs and outputs, making it gradually become the preferred solution for many applications.

\begin{figure}[ht]
    \centering
    \includegraphics[width=\linewidth]{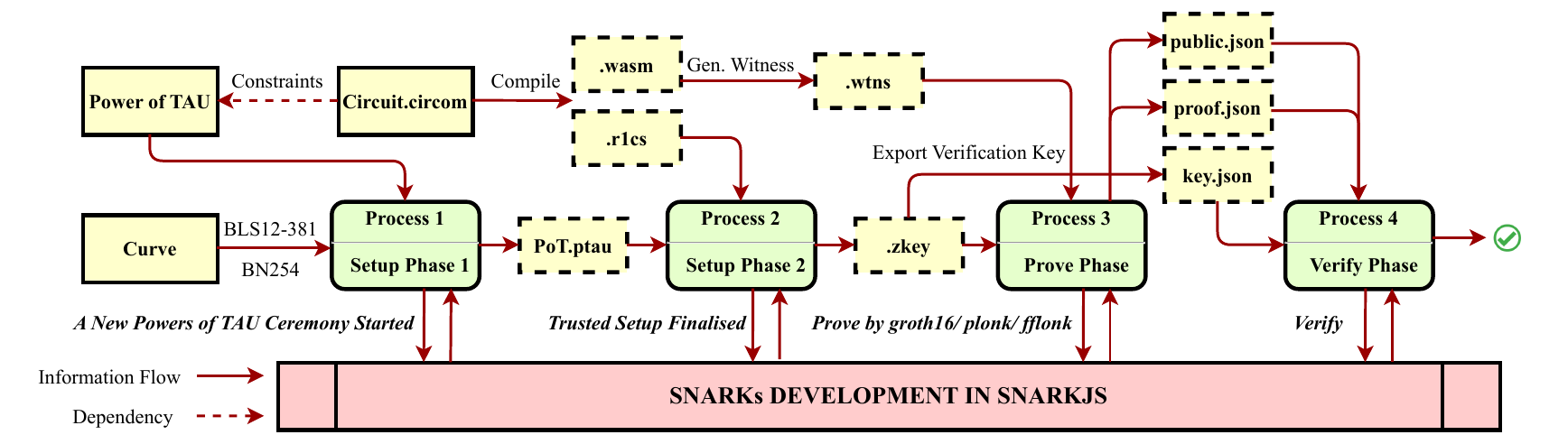}
    \caption{Implementing ZK-SNARKs protocols in \ac{WASM} and JavaScript (\emph{snarkjs})~\cite{snarkjs}. The solid rectangle represents user-defined parameters or circuit templates, while the dashed rectangle represents files generated by \emph{snarkjs}.}
    \label{fig:snarkjs}
\end{figure}

\paragraph{Practical Implementation of SNARKs in \emph{snarkjs}} \autoref{fig:snarkjs} illustrates the pratical implementation workflow of SNARKs in \emph{snarkjs}, divided into four stages:

\begin{enumerate}
    \item \textbf{Setup Phase 1:} The first stage of ZKP setup, corresponding to process 1 in \autoref{fig:snarkjs}, involves a universal trusted setup to generate proving and verification keys, producing \emph{toxic waste} that must be discarded to prevent false proofs and ensure system soundness.~\cite{tau2019} To address this, Koh Wei et al.~\cite{tau2019} proposed the \emph{Perpetual Powers of Tau Ceremony}, a cryptographic ceremony based on \ac{MPC}, for the entire community, alleviating the burden on individual teams. This ensures that as long as one honest participant discards the \emph{toxic waste}, the computations' results remain trustworthy. 
    
    \item \textbf{Setup Phase 2:} Unlike \emph{Plonk} and \emph{Fflonk}, \emph{Groth16} requires not only a universal setup named \emph{Setup Phase 1}, but also an additional circuit-specific trusted setup named \emph{Setup Phase 2} represented in \autoref{fig:snarkjs}, which has a similar process to the former. Both phases of the setup process are based on \ac{MPC} to add an extra source of entropy, which is called \emph{contribution} by ~\cite{baylina2019snarkjs}, to the target files, ensuring the security of the generated files.
    
    \item \textbf{Prove \& Verify:} Each proof system uses its own securely generated proof key \emph{.zkey} and the witness \emph{.wtns} generated after circuit compilation as inputs to produce \emph{proof.json} and \emph{public.json} containing public inputs and outputs, as shown in \autoref{fig:snarkjs}.\footnote{Notably, \emph{snarkjs} can generate the verifier smart contract in \emph{Solidity}, facilitating developers' rapid deployment and verification of SNARKs on-chain.}

\end{enumerate}

\subsection{ZK-Friendly Hash Function} \label{subsubsec: hash}
As discussed in \autoref{subsec: merkletree}, the algorithms used in the commonly employed Merkle tree data structure in blockchain rely heavily on hash functions. The performance of these algorithms depends partly on the tree's depth and largely on the efficiency of the hash function used. Although traditional and well-established hash functions like \emph{SHA256} can run quickly on modern CPUs, they operate at the bit level, which differs significantly from the large prime field elements used in modern ZK proof systems, such as 256-bit in \emph{Halo2}~\cite{halo2kzg} and 64-bit in \emph{Plonky2}~\cite{Plonky2}~\cite{taceoblog}. This discrepancy means that these hash functions cannot be naturally adapted for efficient operation within these systems. Consequently, modern ZK-friendly hash functions that operate directly on prime fields have emerged. Roman~\cite{taceoblog} categorizes these into three types, as shown in \autoref{tab:hash}. 

\begin{table*}[tb]
    \centering
    \resizebox{\textwidth}{!}{
    \begin{tabular}{ccc|ccc}
    \toprule
        Type 1 & Low-degree & $y = x^d$ & MiMC~\cite{albrecht2016mimc} GMiMC~\cite{ben2020} & Neptune~\cite{alan2021} & Poseidon~\cite{grassi2021p1} Poseidon2~\cite{grassi2023} \\
    \midrule
        Type 2 & Low-degree equivalence & $y = x^{\frac{1}{d}} \implies x = y^d$ & Rescue~\cite{aly2019} & Griffin~\cite{grass2022} & Anemoi~\cite{bouvier2022} \\
    \midrule
        Type 3 & Lookup tables & $y = T[x]$ & Reinforced Concrete~\cite{grassi2022rc} & Tip5~\cite{alan2023} & Monolith~\cite{grass2023} \\
    \bottomrule
    \end{tabular}
    }
    \caption{Selected popular ZK-friendly hash functions based on three different design approaches.}
    \label{tab:hash}
\end{table*}

Since \emph{MiMC}~\cite{albrecht2016mimc}, a series of hash functions based on low-degree round functions have been introduced, such as \emph{GMiMC}~\cite{ben2020} and Poseidon~\cite{grassi2021p1}, which can be considered the first generation of ZK hashes. These functions require many rounds for security but involve fewer multiplications per round, offering better performance than bit-based hash functions like \emph{SHA-2} and are relatively easy to analyze for cryptanalytic attacks. Like the first, which is most commonly constructed from the power map $y = x^d$, the second type uses an equivalent representation $y = x^{\frac{1}{d}}$, which, considering the prover's performance, improves the algorithm's runtime in ZK circuits at the expense of computation speed in modern CPUs, such as \emph{Griffin}~\cite{grass2022}, \emph{Rescue}~\cite{aly2019}. A recent class of hash functions, the third type, uses lookup tables $y = T[x]$ for efficient computations in modern ZK-proof systems. This method decomposes large prime field elements into smaller ones, performs a table lookup, and reassembles the result. Advantages include efficient table lookups and strong security with fewer rounds. However, the proof system must support lookup arguments. Notable designs include \emph{Reinforced Concrete}~\cite{grassi2022rc}, optimized for 256-bit prime fields; \emph{Tip5}~\cite{alan2023}, for a 64-bit prime field; and \emph{Monolith}~\cite{grass2023}, which achieves \emph{SHA-3}-level performance for multiple field sizes.

This paper focuses on the implementation of the first type of hash function and benchmarks it across multiple SNARKs supported by \emph{snarkjs}. This focus is due to two reasons: first, not all proof systems support the third type of hash function that relies on lookup tables; second, the second type of hash function, while equivalent in design to the first, improves on-circuit performance at the cost of off-circuit efficiency. Therefore, we evaluate the performance differences among the first type of hash functions, leaving the assessment of the other two types for future work.

\section{Benchmark Results} \label{bench}
Our experiments focused on the \emph{circom}-\emph{snarkjs} toolkit, aiming to provide benchmarking and comparative analysis of the performance of ZK-friendly hash functions and SNARKs protocols. We have released this implementation as open source~\cite{hanze24zklib}. Specifically, we used \emph{circom} to write frontend circuits implementing the logic of the hash functions and employed \emph{snarkjs} to perform zero-knowledge proofs using three backend proof systems: \emph{Groth16}, \emph{Plonk}, and \emph{Fflonk}.\footnote{All test data results in the setup process exclude the setup phase 1 described in \autoref{subsec: zkp}. We use the runtime for generating the \emph{.zkey} file in phase 2 as the runtime metric for the setup stage.} \emph{snarkjs} supports two curves, \emph{BLS12-381} and \emph{BN254}, and we focused on the latter for its efficient computational performance. All measurements were conducted on a MacBook Pro M1.

\subsection{Test Vector} \label{subsec: tv}
For hash functions, we selected five widely used and ZK-friendly low-degree round functions defined in \autoref{subsubsec: hash}: \emph{MiMC}~\cite{albrecht2016mimc}, \emph{GMiMC}~\cite{albrecht2019}, \emph{Neptune}~\cite{neptune2021}, \emph{Poseidon}~\cite{grassi2021p1}, and \emph{Poseidon2}~\cite{grass2023}. Among these, we implemented \emph{GMiMC}, \emph{Neptune}, and \emph{Poseidon2} in \emph{circom},\footnote{We set \emph{GMiMC} with 226 rounds of round constants, \emph{Neptune} with 6 full rounds and 68 partial rounds, and \emph{Poseidon2} with 8 full rounds and 56 partial rounds.}  while using templates from \emph{circom}lib for \emph{MiMC} and \emph{Poseidon} circuits. For the proof system, we covered three SNARKs protocols: \emph{Groth16}, \emph{Plonk}, and \emph{Fflonk}.

On this basis, we chose to construct a common cryptographic data structure, the Merkle tree, as defined in \autoref{subsec: merkletree} \autoref{equation1} to create test vectors. By building Merkle trees of different depths, we obtained test circuits with varying levels of constraints to use as experimental samples. We measured setup, prove, and verify stages, recording and comparing the runtime and RAM consumption across the five hash functions and three SNARKs protocols.

\begin{figure}[tb]
    \begin{subfigure}[t]{0.495\textwidth}
    \includegraphics[height=0.141\textheight, trim=0 0 147mm 0, clip]{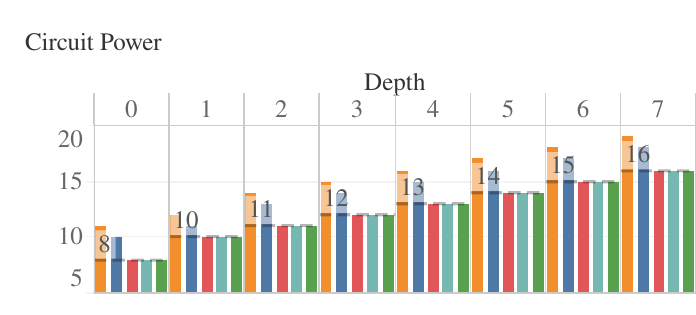}
    \vskip -0.2cm
    \caption{For the R1CS-based (\emph{groth16})}
    \label{fig:c1}
    \end{subfigure}
    \hfill
    \begin{subfigure}[t]{0.495\textwidth}
    \includegraphics[height=0.141\textheight, trim=121mm 0 0 0, clip]{figures/constraints.pdf}
    \vskip -0.2cm
    \caption{For the Plonkish-based (\emph{plonk}, \emph{fflonk})}
    \label{fig:c2}
    \end{subfigure}
    
\caption{\ref{fig:c1} and \ref{fig:c2} respectively show the relationship between the circuit power of templates containing five types of ZK-friendly hash functions and the depth of the Merkle tree in R1CS-based proof systems (\emph{groth16}) and Plonkish-based proof systems (\emph{plonk}, \emph{fflonk}). In each template for different Merkle tree depths, the most efficient hash function's circuit power is highlighted.}
\Description{}
\label{fig: constraints}
\end{figure}

\subsection{Circuit Power} \label{subsec: circuit power}
Due to differences in SNARKs protocol arithmetization as defined in \autoref{subsec: zkp}, the two arithmetic proof systems based on Plonkish and R1CS differ in the form and number of constraints after circuit compilation. \emph{snarkjs} uses \emph{circuit power} and \emph{Plonk circuit power}\footnote{For convenience, unless otherwise stated, circuit power will refer to the circuit size under all three proof systems.} to represent the size of the circuits defined by the number of constraints for \emph{Groth16} based on R1CS, and \emph{Plonk} and \emph{Fflonk} based on Plonkish. The unit of measurement for circuit power is the number of constraints, representing the scale of constraints obtained under a certain arithmetic form. For example, with circuit power = 10, the constraints range is between $2^9$ and $2^{10}$. The measurement of circuit power is directly related to prover performance and is particularly important for SNARKs during the setup phase. Circuit power determines the size of the trusted files generated during the setup phase 1-\emph{Perpetual Powers of Tau Ceremony} as defined in \autoref{subsec: zkp}, requiring the power of tau to match it. Therefore, the larger the circuit power, the larger the power of tau, and the greater the computational resources consumed during the setup phase under the same proof system.

Our observations in \autoref{fig: constraints} revealed that \emph{Neptune}, \emph{Poseidon}, and \emph{Poseidon2} have the least circuit power under R1CS for \emph{Groth16}, while \emph{MiMC} has the least circuit power under \emph{Plonk} and \emph{Fflonk} for Plonkish. This result is consistent with the descriptions in \autoref{subsec: zkp} regarding the two types of arithmetization, where R1CS focuses on multiplication gates and thus has more concise constraints compared to Plonkish. Moreover, this result largely depends on the parameter settings of the hash functions, such as the constraints' scale increasing with more rounds of permutation for the same hash algorithm. Additionally, since Plonkish supports multiple logic gates beyond just multiplication and addition, there may be room for optimization in our implementation for \emph{Plonk} and \emph{Fflonk}. The results regarding circuit power and subsequent benchmarks only reflect the evaluation under the parameter settings defined in \autoref{subsec: tv} in our current implementation.

\subsection{Hash Functions}

\begin{figure}[ht]
    \centering
    \begin{subfigure}{0.5\linewidth}
        \centering
        \includegraphics[width=\linewidth]{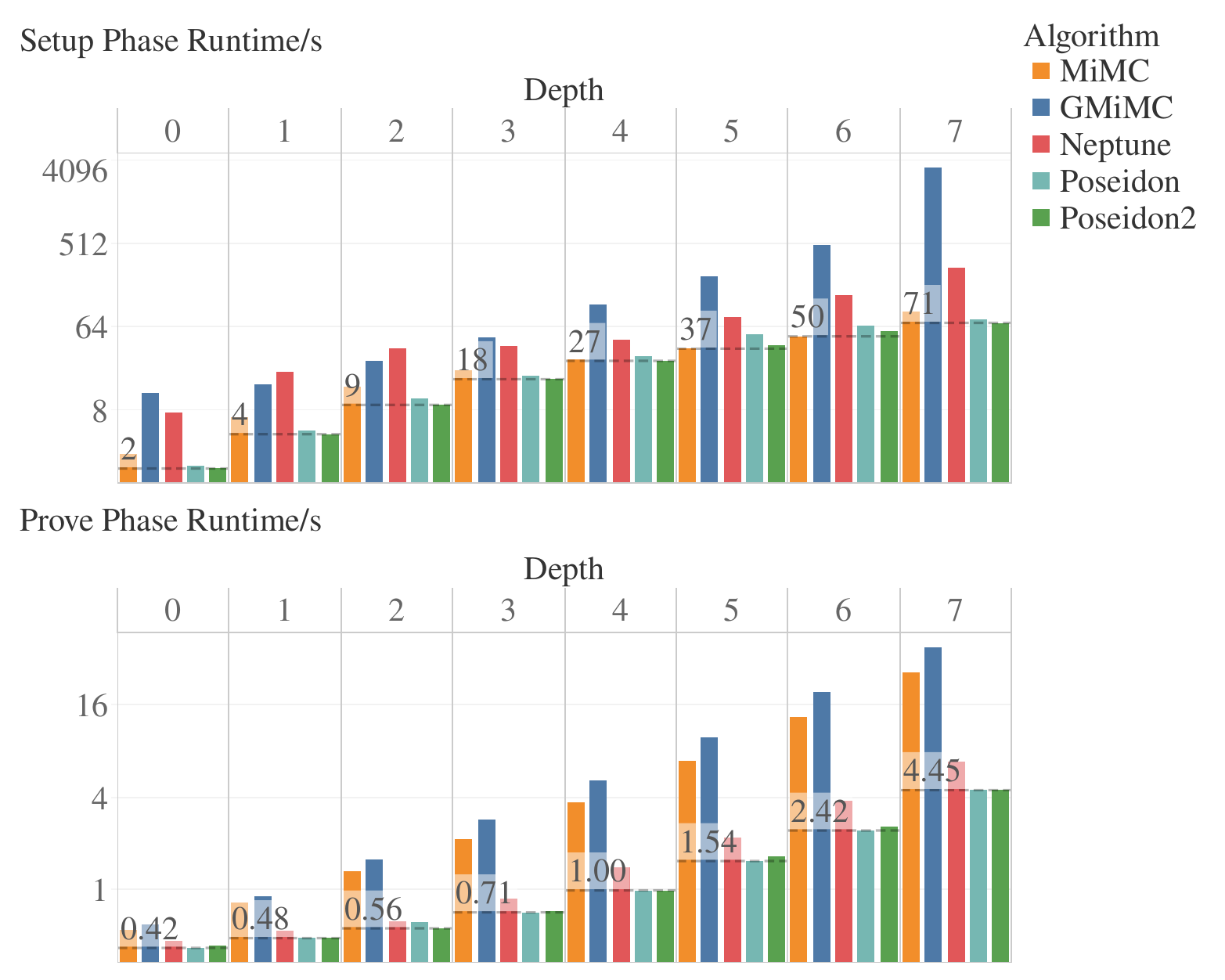}
        \caption{Runtime}
        \label{fig:run_hash}
    \end{subfigure}%
    \begin{subfigure}{0.5\linewidth}
        \centering
        \includegraphics[width=\linewidth]{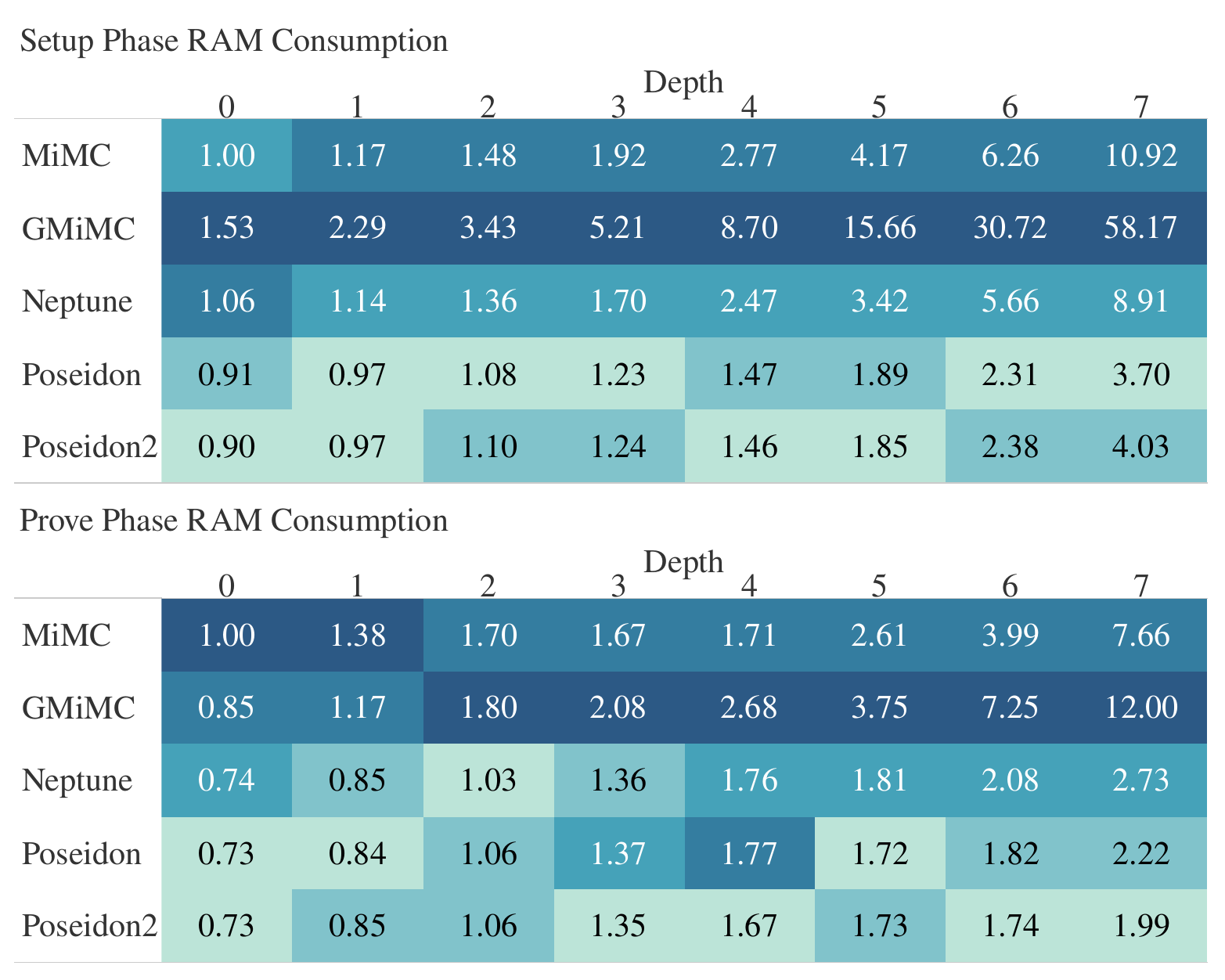}
        \caption{RAM Consumption}
        \label{fig:ram_hash}
    \end{subfigure}
    \caption{Figure \ref{fig:run_hash} shows the relationship between the runtime of the setup and prove phases in the \emph{groth16} proof system and the depth of the Merkle tree for circuit templates containing five types of ZK-friendly hash functions. In each template for different Merkle tree depths, the runtime of the most efficient hash function is highlighted. Figure \ref{fig:ram_hash} illustrates the ratio of memory consumption among these different hash function circuit templates at various Merkle tree depths. The memory consumption of a single MiMC at depth 0 is set as the baseline value of unit 1. In each pane, resource-intensive tests are shown in dark colors, while efficient ones are in light colors.}
    \label{fig: hash}
\end{figure}

\paragraph{Setup} For the setup phase, we found that the \emph{Poseidon} and \emph{Poseidon2} functions were the most efficient in both runtime and RAM consumption, as shown in \autoref{fig: hash}. Specifically, on the largest test circuit, i.e., constructing a Merkle tree with a depth of 7, the runtimes for \emph{Poseidon} and \emph{Poseidon2} remained around 70 seconds, significantly less than \emph{GMiMC}'s 3,476 seconds and 284 seconds. Additionally, their RAM consumption was over 50\% lower than that of all other tested functions.

\paragraph{Prove} According to \autoref{fig:run_hash}, \emph{Poseidon} and \emph{Poseidon2} functions maintained the highest efficiency in runtime as well. For instance, in the circuit with a Merkle tree depth of 7, both functions completed the proof for 127 $(=2^7-1)$ hash operations within 4.5 seconds. In terms of RAM consumption, besides \emph{Poseidon} and \emph{Poseidon2}, \emph{Neptune} also demonstrated good performance. In the circuit with a depth of 7, \emph{Neptune}'s RAM consumption was slightly higher than \emph{Poseidon}'s but still significantly lower than \emph{MiMC} and \emph{GMiMC}, reducing RAM consumption by over 60\% and 75\%, respectively.

To better illustrate the memory consumption in the prove and setup phases, \autoref{fig:ram_hash} shows the ratio of RAM consumption between the prove and setup phases for different test circuits. The results indicate that, except for \emph{GMiMC}, where the prove phase consistently consumed less RAM than the setup phase, there is no consistent pattern in the RAM consumption relationship between these two phases for the other four hash functions. The RAM consumption in the prove and setup phases varied widely.

\paragraph{Verify} We also recorded the RAM consumption for the verify phase in detail, showing that it was much lower than in the prove and setup phases. Therefore, we did not visualize or further analyze it. For runtime, all test cases exhibited an average runtime of around 400ms during the verify phase, consistent with the expected performance characteristics of SNARKs protocols.

\subsection{Proof Systems} \label{subsec: pf}

\begin{figure}[ht]
    \centering
    \begin{subfigure}{0.5\linewidth}
        \centering
        \includegraphics[width=\linewidth]{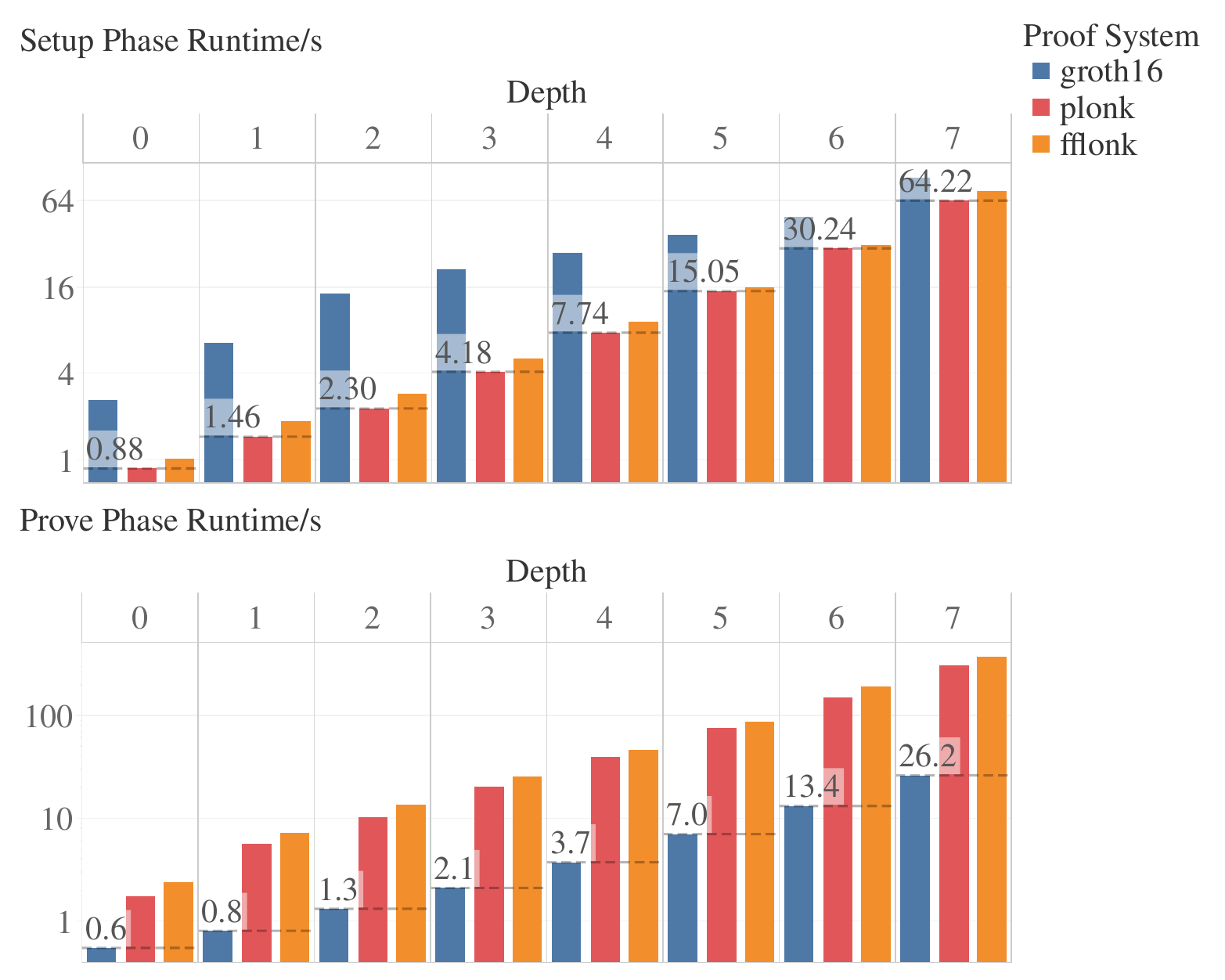}
        \caption{Runtime}
        \label{fig:run_proof}
    \end{subfigure}%
    \begin{subfigure}{0.5\linewidth}
        \centering
        \includegraphics[width=\linewidth]{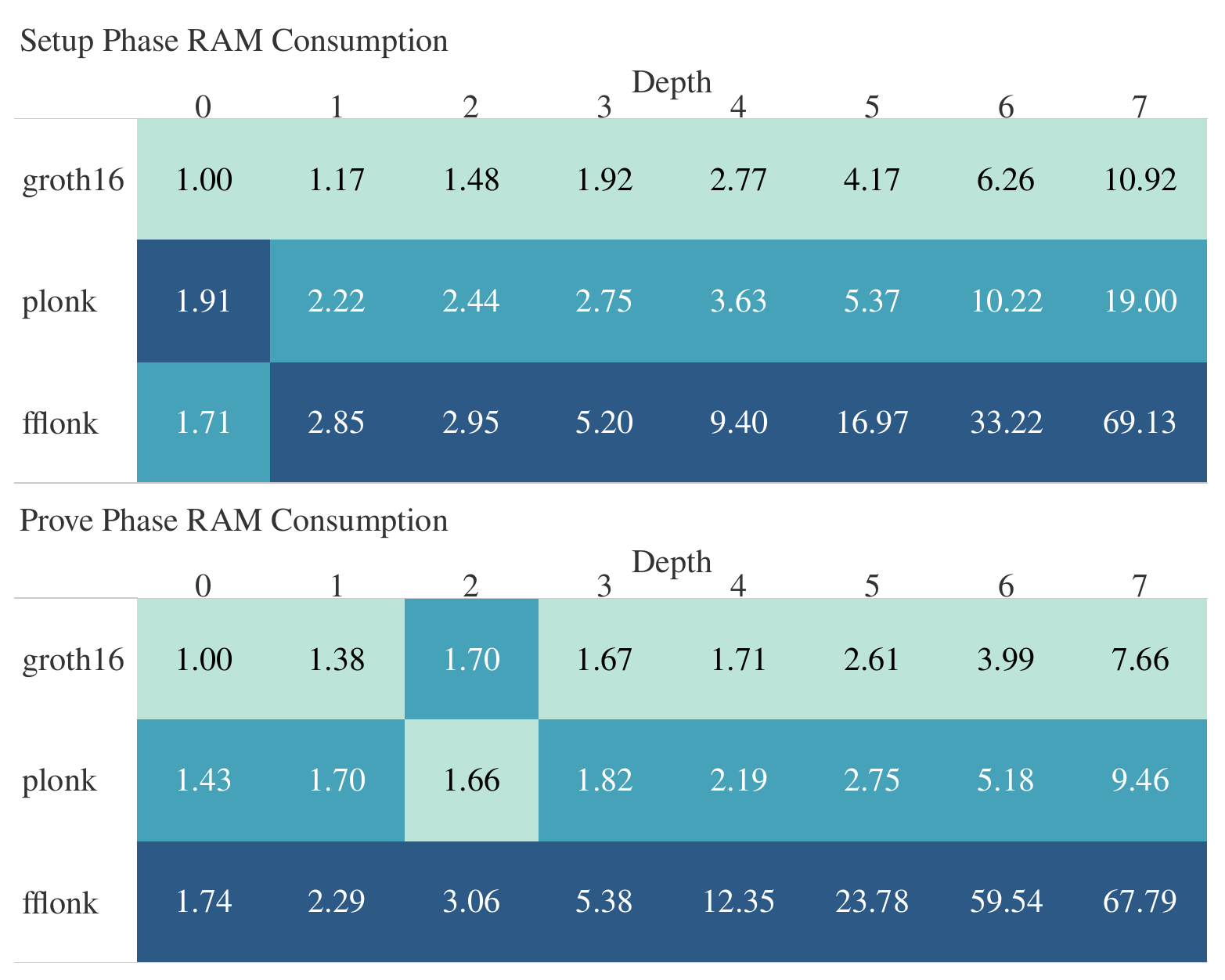}
        \caption{RAM Consumption}
        \label{fig:ram_proof}
    \end{subfigure}
    \caption{Figure \ref{fig:run_proof} shows the relationship between the runtime of the setup and prove phases and the depth of the Merkle tree for \emph{MiMC} hash function circuit templates across three proof systems. In each template for different Merkle tree depths, the runtime of the most efficient hash function is highlighted. Figure \ref{fig:ram_proof} illustrates the ratio of memory consumption among these test circuits running on different proof systems at various Merkle tree depths. The memory consumption of a single \emph{MiMC} in \emph{groth16} at depth 0 is set as the baseline value of unit 1.}
    \label{fig: proof}
\end{figure}

\paragraph{Setup} We found that during the setup phase, \emph{Groth16} required the most runtime,\footnote{In practical use, for security considerations, \emph{Groth16} requires adding sufficient rounds of entropy in setup phase 2 to obtain a final trusted \emph{.zkey} file, unlike \emph{Plonk} and \emph{Fflonk}. Even so, under the experimental conditions in this paper, \emph{Groth16} still shows the lowest efficiency in terms of runtime during the setup stage.} while \emph{Plonk} was the most efficient, as shown in \autoref{fig:run_proof}. For instance, in the test circuit with a depth of 7, \emph{Plonk} and \emph{Fflonk} had runtimes of approximately 60s and 70s, respectively, whereas \emph{Groth16} exceeded 90s. However, in terms of RAM consumption, \emph{Fflonk} required the most memory, while \emph{Groth16} required the least, as shown in \autoref{fig:ram_proof}. When processing the same \emph{MiMC}-based test circuits, the memory consumption for \emph{Plonk} and \emph{Fflonk} was higher than for \emph{Groth16}. This is due to the differences in constraints within \emph{Plonk}-type circuits, as discussed in \autoref{subsec: circuit power}. For example, in a test circuit with a Merkle tree depth of 0 (containing only 1 \emph{MiMC} hash operation), \emph{Plonk} constraints exceed regular constraints by 444. Additionally, \emph{Fflonk}’s memory consumption was significantly higher than \emph{Plonk}’s. Particularly, \emph{Fflonk} requires a \emph{PTAU} file power for the TAU Ceremony that is at least three orders of magnitude larger than \emph{Plonk}'s for the same \emph{Plonk} circuit power. For instance, in a test circuit with a Merkle tree depth of 7 (\emph{Plonk} circuit power of 19), \emph{Plonk} requires a \emph{PTAU} file power of 19, whereas \emph{Fflonk} requires an power of 22.

\begin{figure}[tb]
    \begin{subfigure}[t]{0.495\textwidth}
    \includegraphics[height=0.141\textheight, trim=0 0 130mm 0, clip]{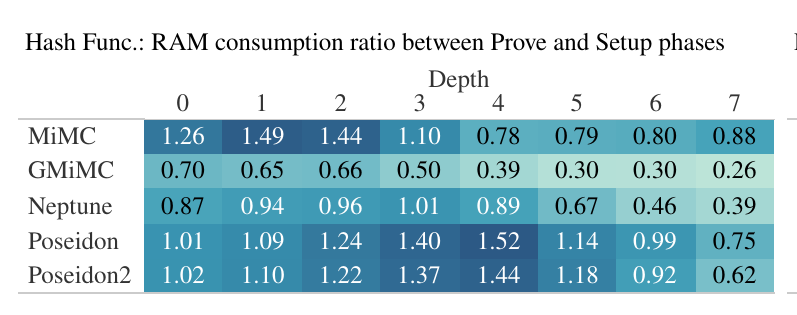}
    \vskip -0.2cm
    \caption{Across Hash Functions}
    \label{fig:r1}
    \end{subfigure}
    \hfill
    \begin{subfigure}[t]{0.495\textwidth}
    \includegraphics[height=0.141\textheight, trim=132mm 0 0 0, clip]{figures/R.pdf}
    \vskip -0.2cm
    \caption{Across Proof Systems}
    \label{fig:r2}
    \end{subfigure}
    
\caption{\ref{fig:r1} shows the ratio of RAM consumption during the prove phase to the RAM consumption during the setup phase for circuit templates containing five types of ZK-friendly hash functions in \emph{groth16} at different Merkle tree depths. \ref{fig:r2} shows the same ratio for MiMC hash function circuit templates across three proof systems at different Merkle tree depths. In each pane, resource-intensive tests are shown in dark colors, while efficient ones are in light colors.}
\Description{}
\label{fig: R}
\end{figure}

\paragraph{Prove} In the prove phase, \emph{Groth16} demonstrated efficiency in both runtime and RAM consumption. As shown in \autoref{fig:run_proof}, \emph{Groth16} outperformed in prover time and RAM consumption. Furthermore, a unique characteristic of SNARKs is the fixed size of the generated ZKP; \emph{Groth16}’s ZKP is approximately 800 bytes, smaller than \emph{Plonk}’s and \emph{Fflonk}’s 2kb. To more accurately reflect RAM consumption during different SNARKs phases, \autoref{fig: R} shows the RAM consumption ratio between the prove and setup phases. This indicates that for \emph{Groth16}, the relationship between the two is not constant, while for \emph{Fflonk}, RAM consumption during the prove phase is likely higher than during the setup phase, and vice versa for \emph{Plonk}.

\paragraph{Verify} In addition to the above conclusion that the verify phase runtime stabilizes at approximately 400ms, we further compared the gas consumption for deploying and calling the verifier smart contracts (generated by \emph{snarkjs}) on different blockchain environments (\emph{Ethereum} and \emph{Hedera}).\footnote{Since the default \emph{verifier.sol} generated by \emph{snarkjs} uses a view function for verification, it does not consume gas as it does not modify the on-chain state. To address this, we added events to record verification results, which changes the function type and allows us to measure the gas consumption.} For contract deployment, \emph{Groth16} consumed the least gas, whether on Ethereum or Hedera, at approximately 458k and 83k, respectively.\footnote{\emph{Hedera}'s gas consumption measurement is mostly identical to \emph{Ethereum}'s, with different handling only for storage-related opcodes. \emph{Ethereum} imposes a fixed gas consumption regardless of storage duration, while \emph{Hedera} considers storage duration, measuring gas consumption for storage operations, like \emph{LOG} and \emph{SSTORE}, in a rental-like manner. This explains the different gas consumption observed for contract deployment across the two networks.} This saved over 91\% and 80\% compared to the most gas-consuming \emph{Fflonk}'s 5,299k and 434k, respectively. For calling the verification function, \emph{Fflonk} consumed the least gas at approximately 209k, saving approximately 5\% compared to \emph{Groth16}’s 219k and nearly 30\% compared to \emph{Plonk}’s 298k. This result is consistent with the expectations in \autoref{subsec: zkp}, where \emph{Fflonk} trades off computational resources in the setup and prove phases for reduced on-chain verification costs.
\section{A Cost-Efficiency Exploration on SNARKs} \label{exploration}
To explore the cost-efficiency potential of SNARKs, we have applied the concept of batching to ZKP-based privacy protection protocols. Although we use \emph{TC}, one of the most active privacy-preserving protocols on \emph{Ethereum} and implemented using the \emph{circom-snarkjs} tool, as our baseline for clearer illustration, our aim is to extend its application to broader fields beyond cryptocurrency transactions, such as privacy protection in healthcare data. By adding the role of a sequencer to batch process user transaction requests, we aim to achieve cost savings and increased efficiency. Additionally, the introduction of a sequencer, which maintains a list of non-banned addresses, brings a certain level of elastic censorship resistance to the ZKP mixer. Before delving into the practical effectiveness of this approach, this section outlines the methods for improving ZKP mixers.

\begin{figure}[ht]
    \centering
    \includegraphics[width=\linewidth]{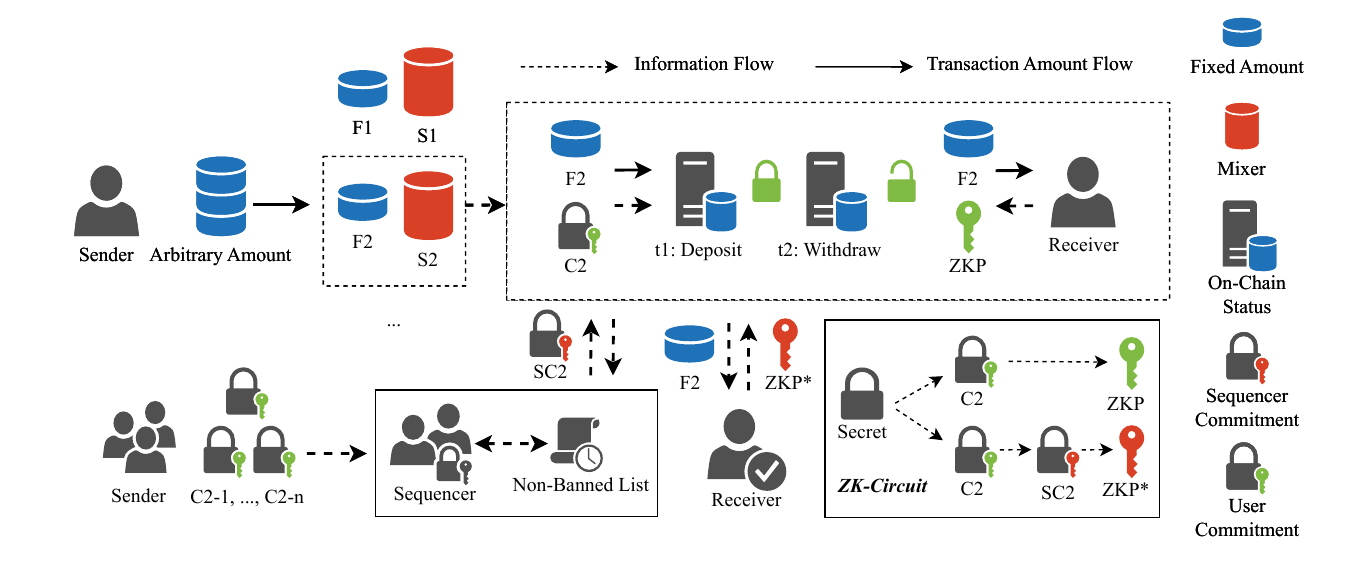}
    \caption{The workflow in \emph{TC} and in the protocol after adopting SNARKs for batch processing.}
    \Description{}
    \label{fig: flow}
\end{figure}

\subsection{Overview} \label{subsec: overview}
To simplify the protocol architecture, we categorize participants into two main roles: users and sequencers, who both act as ZKP provers. Users come from any EVM-compatible blockchains and possess either native assets or \emph{ERC-20} compliant tokens. The sequencer role is performed by a network of nodes that meet the security assumptions detailed in \autoref{subsec: threat}.

In \emph{TC}, after a sender initiates a privacy-protected transaction of any amount, the amount is split into fixed units and deposited into corresponding mixer pools. For example, in \autoref{fig: flow}, the amount is divided into three fixed-amount transactions and sent to mixer pools $S1$, $S2$, and $S3$. After allocation, \autoref{fig: flow} uses a single mixer pool, $S2$, to explain the protocol's workflow. Users must secretly generate a commitment $C2$ represented as \(H_{\it Pedersen}: (\it secret, \it nullifier) \rightarrow C2\). The definitions of $\it secret$ and $\it nullifier$ follow \emph{TC}'s specifications~\cite{TornadoCash}. Subsequently, users submit $C2$ along with the $F2$ transaction amount to \emph{TC}'s smart contract, which records the commitment in an incremental Merkle tree by running \autoref{equation2} defined in \autoref{subsec: merkletree}. To withdraw, the receiver must reproduce this process in ZK circuits to obtain a ZKP \(\pi_{\it user}\), then submit \(\pi_{\it user}\) to the smart contract to prove they know \emph{secret} needed to generate the commitment, thus verifying their identity by running \autoref{equation3} defined in \autoref{subsec: merkletree}.

In our method, users do not need to submit commitments sequentially to the on-chain smart contract. Instead, the sequencer batches them using \autoref{equation1} from \autoref{subsec: merkletree} and generate the hash of the Merkle tree root $SC2$. The sequencer then submits $SC2$ to the on-chain smart contract, replacing the individual user submissions. To ensure the correctness of the sequencer’s batch processing, the sequencer generates a ZKP \(\pi_{\it sequencer}\) for the Merkle tree creation process for each batch and submits \(\pi_{\it sequencer}\) along with $SC2$ to the on-chain smart contract. The details for verifying identity during the withdrawal process are specifically elaborated in the following \autoref{subsec: reducecost}.

By using this method, assuming a batch size of 32, each 32 commit operations required for 32 privacy-protected transactions in \emph{TC} are reduced to 1. Since commit operations modify the state of the on-chain smart contract, this method significantly reduces the cost of on-chain interactions. The specific results of this cost reduction are described in \autoref{subsec: reducecost}.

\subsection{Threat Model} \label{subsec: threat}
We provide a detailed analysis of the threat model for our solution, particularly explaining the construction of the sequencer below.  

\subsubsection{Security Goals and Trust Assumptions}
Sequencers, as the connectors of the entire protocol, play a crucial role in the overall security of the protocol. They should be composed of a set of trusted nodes and need to meet the following two security properties:
\begin{itemize}
    \item \textbf{Liveness:} Sequencers must promptly package and submit commitments after a required number of privacy transactions are successfully deposited.
    \item \textbf{Correctness:} Each ZKP generated by the sequencer and the submitted commitment must consistently adhere to the rules and include all successfully deposited transactions.
\end{itemize}

Based on the above security objectives, we present the trust assumptions in Theorem 1:

\begin{globaltheorem}
The privacy protection solution based on sequencer batching meets the security objectives when the following three security assumptions are satisfied:

\begin{itemize}
    \item The utilized ZK-SNARK protocol is robust.
    \item At least one honest node exists within the sequencers.
    \item The blockchains where the protocol is deployed are consistent and live.
\end{itemize}

\end{globaltheorem}

\subsubsection{Potential Attack Vectors and Mitigation} \label{subsubsec: attacks}
Based on motivations, we classify adversaries aiming to maximize profit as \emph{rational adversaries}, and those willing to compromise system security at their own expense as \emph{non-rational adversaries}~\cite{majority2018}. Based on permissions and capabilities, adversaries are categorized as \emph{administrators} and \emph{non-administrators}. \emph{Non-administrators} have permissions equivalent to regular users, limited to deposit and withdrawal operations via smart contract interfaces within mixer pools. \emph{Administrators}, with sequencer-equivalent permissions, directly manage deposits within smart contracts.

The sequencers in our approach take on responsibilities similar to those in ZK Layer2~\cite{zksyncera, Starknet}, including transaction sorting, packaging, and generating corresponding ZKPs, and collectively maintaining a Non-banned address list to achieve elastic censorship resistance. Based on this, we refer to Motepalli et al.'s research~\cite{soksequencers2023} on decentralized sequencers and provide a design example of its main modules in the following cases:

\paragraph{Front-Running Attacks.} \emph{Rational adversaries} intercept withdrawal requests from honest users, alter the recipient address, and use the ZKP to steal deposits. A defense, as used in \emph{TC}~\cite{TornadoCash}, incorporates critical information, like receipt addresses, into the ZKP proof chain to ensure consistency checks in legitimate withdrawals. This method leverages the soundness property (\autoref{subsec: zkp}) of ZKP , preventing attackers from generating a new ZKP with an altered receipt address.

\paragraph{Replay Attacks.} \emph{Rational adversaries} exploit ZKPs to withdraw funds multiple times after a deposit and withdrawal operation. Our defense requires users to include a random number \emph{nullifier} in the commitment generation process and check the \emph{nullifier}'s hash in the withdrawal smart contract to prevent reuse of ZKPs~\cite{TornadoCash}. This method also leverages the soundness property of ZKP, preventing attackers from altering the \emph{nullifier} in a consumed ZKP to create a forged ZKP and steal assets.

\paragraph{Collusion and Censorship.} A \emph{rational administrator adversary} bribes sequencer nodes to manipulate mixer pool funds or fabricate transactions for profit. Conversely, a \emph{non-rational administrator adversary} inspects transactions and refuses to generate correct commitments, compromising system integrity. An effective defense is to use a \ac{FCFS} mechanism for transaction sorting, delegating this to blockchain validators. For instance, users deposit funds into smart contracts, triggering deposit events validated by validators. Sequencers then sort and package transactions based on FCFS principles, simplifying consensus. Additionally, MPC in smart contracts control fund usage, ensuring fund safety in the mixer pool assuming at least one honest sequencer node.

\paragraph{\ac{DoS}.} A \emph{non-rational administrator adversary}, acting as a transaction packager, refuses to perform corresponding operations, disrupting system liveness. A mitigation measure is to employ specific proposer-selection mechanisms to determine transaction packaging executors for each round, such as round-robin mechanisms~\cite{soksequencers2023}, ensuring timely proposer replacement in the event of a DoS attack.

Ideally, a set of reward and penalty mechanisms should be in place. For example, honest sequencers could receive transaction fee rewards, while dishonest adversaries could face asset forfeiture penalties. The specific design of these mechanisms is left for future work.

\subsection{Reducing On-chain Cost} \label{subsec: reducecost}
In a typical ZKP mixer, users first need to add the secretly generated commitment to the on-chain state managed by the smart contract, to provide a basis for the subsequent withdrawal process of generating ZKP. During this process, invoking the smart contract to complete the on-chain state transition consumes a considerable amount of gas. Referring to \autoref{impl}, in the case where the Merkle tree depth is 20 on-chain, it requires over 900k gas. One of the ideal solutions is to aggregate transactions submitted by users individually off-chain into batches for batch processing, thereby enabling a single on-chain state transition to accommodate commitments from more users. Simultaneously, by proving the off-chain packaging process with ZKP and defining the logic to verify ZKP in the on-chain smart contract, privacy transactions with the same security level can achieve cost savings.

\begin{figure}[ht]
    \centering
    \includegraphics[width=\linewidth]{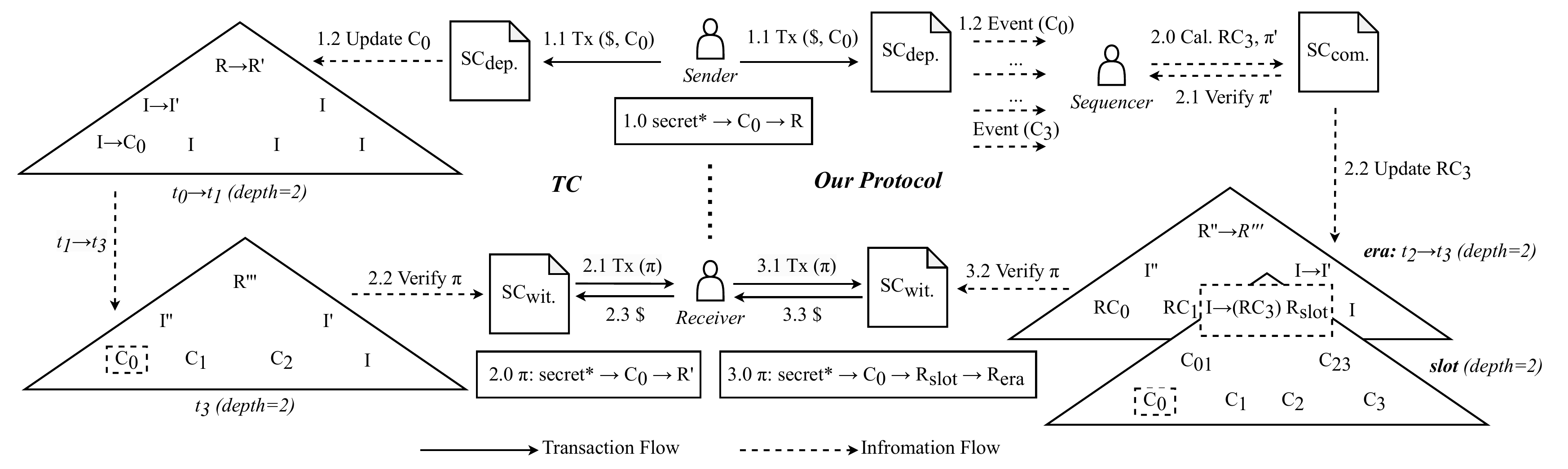}
    \caption{The left and right parts of the diagram respectively illustrate the interaction workflows among the main entities in the \emph{TC} protocol and in our protocol. In \emph{TC}'s workflow, using an incremental Merkle tree of depth 2 as an example, we demonstrate the main steps of completing a \enquote{deposit} at time $t_0$-$t_1$ and a \enquote{withdraw} at time $t_3$. In our solution's, using the same depth 2 incremental Merkle tree \emph{era} and a fully populated Merkle tree \emph{slot} as examples, we show the main steps of the sequencer completing a batch submission at time $t_2$-$t_3$, and the user performing a \enquote{withdraw} at the same time.}
    \label{fig: reducing_cost}
    \Description{}
\end{figure}

As shown in \autoref{fig: reducing_cost}, we designed a combination of two Merkle trees, \emph{era} and \emph{slot}, to achieve this. The \emph{era} tree records on-chain state transitions, with its depth \( d_{\it{era}} \) determining the upper limit of state transitions (\( 2^{d_{\it era}} \)). To maintain protocol continuity, \emph{era} is an incremental Merkle tree, saving on-chain state transition costs when running \autoref{equation2} and \autoref{equation3}. To aggregate off-chain privacy transactions, we use fully populated \emph{slot} trees to store transaction batches. The depth \( d_{\it slot} \) determines the number of transactions aggregated off-chain (\( 2^{d_{\it slot}} \)). At time \( t \), for \( S_i \), each on-chain state transition from \(\it{era}^{t-1}\) to \(\it{era}^{t}\) requires constructing a new \emph{slot}.

Now, let's verify that when using $era$ and \emph{slot}, the proposition for users to prove their authentication is equivalent to that in the baseline protocol. For simplicity, we discuss the scenario where a single user \( U \) conducts a single transaction with a mixer pool \( S \). Their privacy and security parameters are represented as $(\it secret, \it nullifier)$ defined in \autoref{subsec: overview}, and the corresponding commitment is denoted as \( \mathcal{C} \). We define \( S_{\it depositor} \) as the set of users who deposit honestly. \( \emph{slot} \in \emph{era} \) indicates that \emph{slot} is a subtree of \emph{era}. The proposition for users to verify their own authentication in \emph{TC} and our protocol can be expressed separately as Proposition 1 and 2:

\begin{itemize}
    \item \textbf{Proposition 1:} Suppose \( U \) knows a set of \( (\textit{secret}, \textit{nullifier}) \). If the commitment \( \mathcal{C} \) is generated by \( H_{\textit{Pedersen}}(\textit{secret}, \textit{nullifier}) \) and \( \mathcal{C} \) satisfies the verification condition of \autoref{equation3} (i.e., \( \mathcal{C} \in \mathcal{M} \)), then \( U \in S_{\textit{depositor}} \).

    \item \textbf{Proposition 2:} Suppose \( U \) knows a set of \( (\textit{secret}, \textit{nullifier}) \). If the commitment \( \mathcal{C} \) is generated by \( H_{\textit{Pedersen}}(\textit{secret}, \textit{nullifier}) \) and \( \mathcal{C} \) satisfies the verification condition of \autoref{equation3} for \emph{slot}, and \( r_{\textit{slot}} \) satisfies the verification condition of \autoref{equation3} for \emph{era} (i.e., \( \mathcal{C} \in \mathbf{L}_{\it slot} \) and \( r_{\it slot} \in \mathbf{L}_{\it era} \)), then \( \mathcal{C} \in \mathbf{L}_{\it era} \). In this case, \( U \) is considered to be in \( S_{\textit{depositor}} \).
\end{itemize} 

It is easy to observe that in Proposition 2, for \emph{era}, there exists a matrix \emph{slot} of length \( i \) and width \( 2^{d_{\it era}} \) as its subtree, and any \emph{era} can be viewed as a Merkle tree with depth \( d_{\it slot} + d_{\it era} \). This equivalent relationship is stated in Theorem 2 below.

\begin{globaltheorem}
When \( r_{\it slot} \in \mathbf{L}_{\it era} \), any \( \mathcal{C} \in \mathbf{L}_{\it slot} \), then \( \mathcal{C} \in \mathbf{L}_{\it era} \).
\end{globaltheorem}

Thus, we can conclude that the conclusion of Proposition 2, \( U \in S_{\it depositor} \), holds true, and therefore Proposition 1 is completely equivalent to Proposition 2.

Next, we discuss the differences between $era$ and \emph{slot} compared to the baseline protocol in terms of time and space complexity based on \autoref{subsubsec: notations}. In the withdraw operation, users first need to iteratively generate \( r_{\it slot} \) through \( 2^{d_{\it slot}}-1 \) iterations of $H_{\it MiMC}$ calculations, followed by generating \( r_{\it era} \) through \( d_{\it era} \) iterations of $H_{\it MiMC}$ calculations. Therefore, the time complexity is represented as \( O(d_{\it era}+2^{d_{\it slot}}) \), which introduces an additional \( O(2^{d_{\it slot}}) \) compared to the baseline protocol. Additionally, in this process, compared to the baseline protocol's space complexity of \( O(d_{\it era}) \), users need to introduce an additional space complexity of \( O(2^{d_{\it slot}}) \) to calculate \( r_{\it slot} \). Thus, the space complexity is represented as \( O(d_{\it era}+2^{d_{\it slot}}) \). Based on these results, we can infer that the additional time and space complexity introduced by our protocol compared to the baseline protocol are both exponential in terms of \( d_{\it slot} \). Therefore, the configuration of \( d_{\it slot} \) has a close relationship with the prover time for computing \( \pi_{\it sequencer} \) and \( \pi_{\it user} \) in the ZKP circuit, as well as the on-chain costs. We discuss this trade-off in \autoref{impl}.

\subsection{Elastic Censorship and Privacy Assessment} \label{subsec: privacy model}
\begin{figure}[ht]
    \centering
    \includegraphics[width=\linewidth]{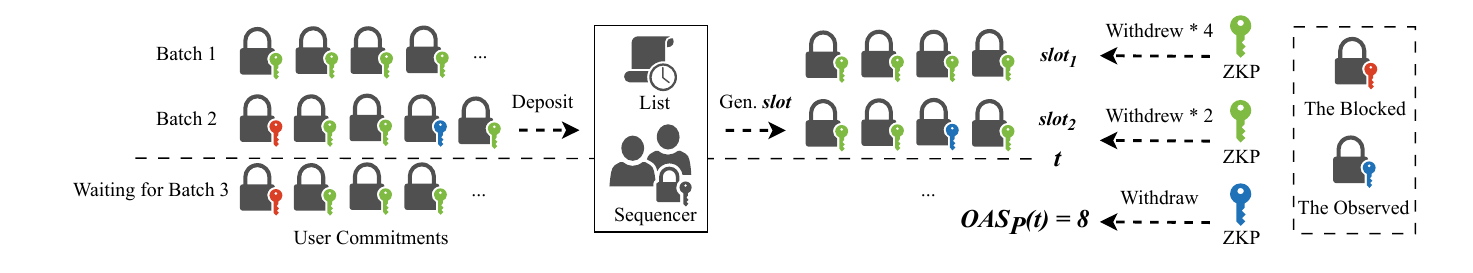}
    \caption{The figure shows that at time $t$, when the sequencer has completed two \emph{slot} with a batch size of 4, $OAS_p(t)$ of an observed \enquote{withdraw} operation is 8, which is the sum of two batch sizes. Although the figure includes some completed \enquote{withdraw} operations, represented as \enquote{Withdrew}, $OAS_p(t)$ of the observed operation is not affected by them.}
    \label{censor}
    \Description{}
\end{figure}

As the enforcer of flexible censorship, the sequencer needs to isolate addresses that are not on the \emph{Non-banned list} during deposit operations. Specifically, following the workflow outlined in \autoref{subsec: overview}, the sequencer will package the corresponding commitments of transactions that have completed on-chain deposits based on the FCFS principle. At this point, the sequencer will package the commitments of addresses that meet the standards according to the list. This ensures that addresses on the blacklist, even if they complete deposits, will not have their transactions packaged into \emph{slot} by the sequencer, and thus will not be able to generate ZKP for withdrawal. In \autoref{censor}, we can see that with a batch size of 4, for example, the sequencer will skip transactions from addresses not on the list and wait for transactions from a sufficient number of legitimate addresses to meet the batch size requirement.

For the privacy assessment, we reference Wang et al.'s privacy model definition for ZKP mixers~\cite{wang2023on}. We only consider the privacy assessment of the observed anonymity set, denoted as $OAS_p(t)$, meaning we assume users do not reveal any privacy during the transaction period, thus preventing the calculation of a simplified anonymity set through on-chain data. Given that a withdraw transaction $Tx_{\it withdraw}$ occurs at a certain time $t$, suppose an adversary attempts to uncover the link between the receiver address $Addr_{\it receiver}$ of this transaction and its corresponding sender address $Addr_{\it sender}$. We will explain the probability $P$ of their success through the following definitions, thereby revealing the privacy of our method.

\begin{itemize}
    \item \textbf{The Observed Anonymity Set ($OAS_p(t)$)}: We define the total number of \emph{slots} constructed by the sequencer at time $t$ as $C_t$, and the depth of each \emph{slot} as $d_{\textit{slot}}$. At this time, the number of completed deposit transactions can be expressed as $d_{\textit{slot}} \cdot C_t$. Assuming that the addresses of each completed deposit transaction are entirely different, the anonymous set of potential $Addr_{\textit{sender}}$ can be represented as $OAS_p(t) = d_{\textit{slot}} \cdot C_t$.
    
    \item \textbf{Privacy Metric ($P$)}: Given $OAS_p(t)$ at time $t$, the probability of correctly linking $Addr_{\textit{receiver}}$ to $Addr_{\textit{sender}}$, assuming that the addresses of each completed deposit transaction are entirely different, can be expressed as $P = \frac{1}{OAS_p(t)}$, i.e., $P = \frac{1}{d_{\textit{slot}} \cdot C_t}$.
\end{itemize}

\autoref{censor} shows the size of $OAS_p(t)$ corresponding to $Tx_{\it withdraw}$ with $C_t = 2$ and $d_{\it slot} = 2$. Observing the expression for privacy $P$, it is evident that with $t$ constant, the more \emph{slot} the sequencer completes within time $t$, and the greater $d_{\it slot}$, the larger $OAS_p(t)$ for the observed withdraw operation. Consequently, the probability of correctly linking the sender and receiver decreases, thereby enhancing privacy.
\section{Implementation and Evaluation} \label{impl}
Our implementation is based on a fork of the \emph{TC} repository, which is released as open source~\cite{hz24zkmixer}. Particularly, we chose the same ZK tools-\emph{snarkjs-circom}, proof system-\emph{Groth16}, elliptic curve-\emph{bn254}, and cryptographic hash function-\emph{\it MiMC} as \emph{TC}.~\cite{TornadoCash} This choice allows for establishing a control group to observe the improvement and trade-offs in protocol performance under a single variable change—namely, modifications in circuit logic. However, in subsequent \autoref{subsec: evaluation}, building upon the conclusions from \autoref{bench}, we incorporate the more efficient \emph{Poseidon2} hash function to optimize the existing protocol and conduct a final comparative analysis.
\subsection{Implementation}
We conducted tests for native token transactions in three deployment environments: \emph{Sepolia}–ETH for \emph{Ethereum}, \emph{BNB testnet}-BNB for \emph{BNB Chain}, and \emph{Hedera testnet}-HBAR for \emph{Hedera}. In this testing environment, we set $d_{\it era} = 20$ and $d_{\it slot} = 5$. At this configuration, the protocol can tolerate a maximum of $2^{20}$ aggregated commitments submitted by the sequencer, with each aggregated commitment containing $2^5$ transactions. The protocol can handle a maximum of $2^{25}$ transactions.

\subsubsection{Circuit} \label{subsubsec: circuit}
In the protocol, there are two main circuits: $\textit{Cir}_{\textit{dep.}}$ and $\textit{Cir}_{\textit{wit.}}$. $\textit{Cir}_{\textit{dep.}}$ is generated and submitted by the sequencer during the deposit operation, while $\textit{Cir}_{\textit{wit.}}$ is generated and submitted by the user during the withdraw operation.

In $\textit{Cir}_{\textit{dep.}}$, the main function is $\textit{Func}_{\textit{slot}}$, which calculates \(r_{\it slot}\) based on an array of leaf nodes with a length of \(2^{d_{\it slot}}\) running by \autoref{equation1}. $\textit{Func}_{\it slot}$ utilizes $\textit{Func}_{\textit{MiMCSponge}}$ template provided by \emph{circomlib}~\cite{circomlib} to implement the process of calculating \(r_{\it slot}\) iteratively using $H_{\it MiMC}$. 

In $\textit{Cir}_{\textit{wit.}}$, there are two main functions: $\textit{Func}_{\textit{hasher}}$ and $\textit{Func}_{\textit{wit.}}$. $\textit{Func}_{\textit{hasher}}$ is responsible for calculating the commitment $\mathcal{C}$, while $\textit{Func}_{\textit{wit.}}$ is the main implementation of $\textit{Cir}_{\textit{wit.}}$.

\begin{enumerate}
    \item \( \textit{Func}_{\textit{hasher}} \) receives a set of secure parameters \( (\it secret, \it nullifier) \) defined in \autoref{subsec: overview} secretly generated by the user as signal inputs and performs \( H_{\it Pedersen}: (\it secret, \it nullifier) \rightarrow \mathcal{C} \). It calls $\textit{Func}_{\textit{Bitfy}}$ and $\textit{Func}_{\textit{Pedersen}}$ two templates provided by \emph{circomlib}.
    \item  $\textit{Func}_{\textit{wit.}}$ receives \((\it secret, \it nullifier)\), \(\pi_{slot}\), \(\pi_{\it era}\), and transaction details such as the receiver address as signal inputs. \((\it secret, \it nullifier)\) is used for authentication, and transaction details is used for security verification, both defined in \autoref{subsubsec: attacks}. \(\pi_{\it slot}\) and \(\pi_{\it era}\) defined in \autoref{subsec: merkletree} compute the path elements and path indices corresponding to \emph{slot} and \emph{era}. Additionally, the signal input includes two root values \(r_{\it era}\) and \(r_{\it slot}\), used for equality comparison after calculating the new root values. $\textit{Cir}_{\textit{wit.}}$ also references a template $\textit{Func}_{\textit{checker}}$ used to verify the existence of leaf nodes by \autoref{equation3} as signal input in the Merkle tree.
\end{enumerate}

\subsubsection{Smart Contract} \label{subsubsec: sc}
Compared to the source code of \emph{TC}, our implementation primarily modifies the abstract smart contract named $SC_{\it protocol}$ that carries the main functions of the protocol. \autoref{lst:protocol} shows the main functions and interfaces of $SC_{\it protocol}$, which inherits from $SC_{\it protocol}$ in \autoref{lst:era}, responsible for updating the on-chain state, i.e., the era incremental Merkle tree. Our modifications to $SC_{\it protocol}$ are specifically reflected in the following three points:
\begin{enumerate}
    \item Compared to the original \emph{TC} which had only one instance of the interface \emph{IVerifier}, we added two instances to verify the ZKPs \(\pi_{\it sequencer}\) and \(\pi_{\it user}\) under Groth16.
    \item The original deposit function $\textit{Func}_{\it dep.}$ was modified so that it no longer triggers an on-chain state update, but only serves as a place for user deposits, and an event $E_{\it dep.}$ was added to record completed deposit transactions for the sequencer to listen to and package.
    \item A function $\it{Func}_{\it com.}$ was added for the sequencer to submit $r_{\it slot}$ after packaging, thus performing the on-chain era state transition. $\it{Func}_{\it com.}$ replaces the original function in \emph{TC} used for user deposits that automatically triggered on-chain state updates, and is responsible for calling \emph{IVerifier} to verify each ZKP \(\pi_{\it sequencer}\) submitted by the sequencer.
\end{enumerate}

\subsection{Evaluation} \label{subsec: evaluation}
For $\textit{Cir}_{\textit{dep.}}$ defined in \autoref{subsubsec: circuit}, the number of constraints $N_{\it dep.}$ in \emph{Groth16}'s R1CS shows an exponential relationship with $d_{\it slot}$: $N_{\it dep.} = 1,320 \times (2^{d_{\it slot}}-1)$. Here, 1,320 represents the number of constraints for a single operation of $H_{\it MiMC}$. Based on the experimental data in \autoref{bench}, we can infer that within the range of $2^{15}$ to $2^{17}$ constraints, the prover time increases linearly. Following this inference, we believe that for $d_{\it slot}$ and $d_{\it era}$ within the ranges of 1 to 9 and 20 to 40 respectively, the prover time of $\textit{Cir}_{\textit{dep.}}$ shows an exponential relationship with the size of $d_{\it slot}$. Similarly, for $\textit{Cir}_{\textit{wit.}}$, the number of constraints $N_{\it wit.}$ shows a linear relationship with $d_{\it slot}$ and $d_{\it era}$: $ N_{\it dep.} = 1,815 + 1,323 \times (2^{d_{\it slot}} + d_{\it era})$. We infer that the prover time of $\textit{Cir}_{\textit{wit.}}$ shows a linear relationship with the size of $d_{\it era}$ and $d_{\it slot}$ within the range of $2^{15}$ to $2^{17}$ constraints.

\begin{figure}[ht]
    \centering
    \begin{subfigure}[b]{0.5\linewidth}
        \centering
        \includegraphics[width=\linewidth]{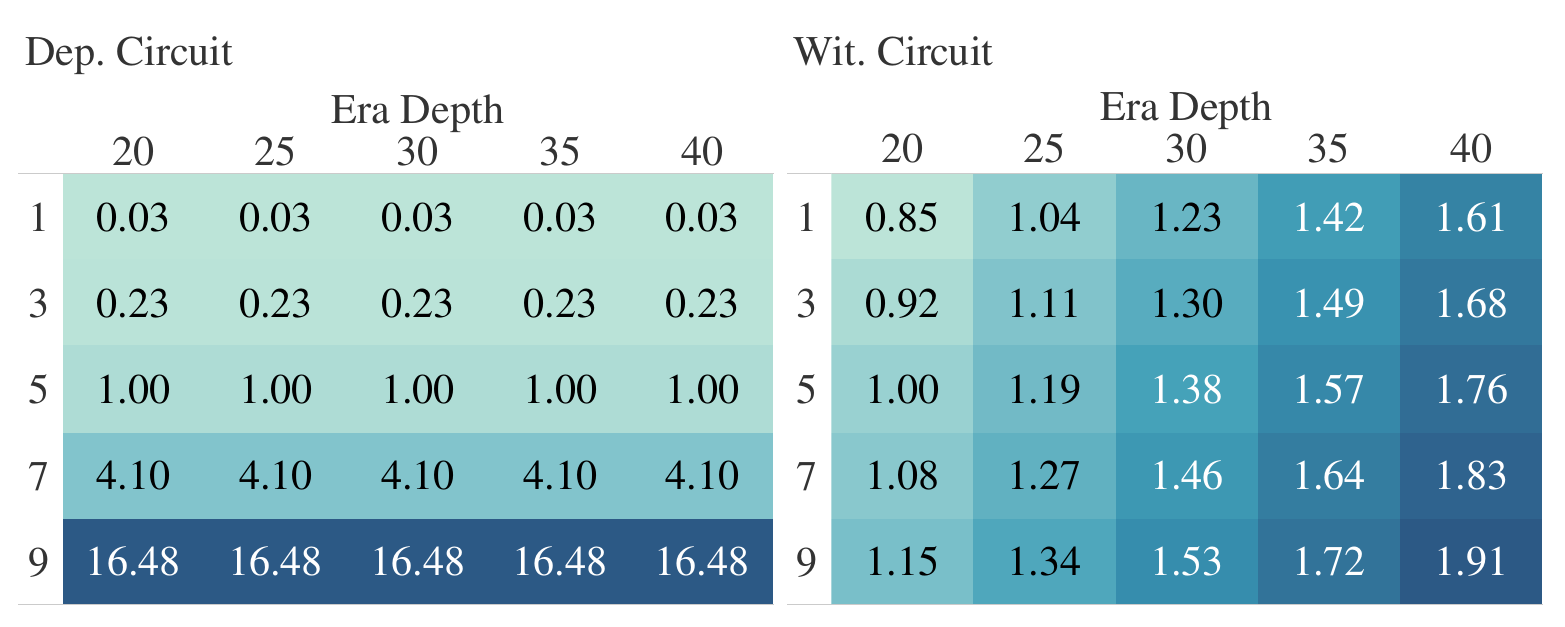}
        \caption{Proof Generation Time}
        \Description{}
        \label{fig:pt_circuit}
    \end{subfigure}%
    \begin{subfigure}[b]{0.5\linewidth}
        \centering
        \includegraphics[width=\linewidth]{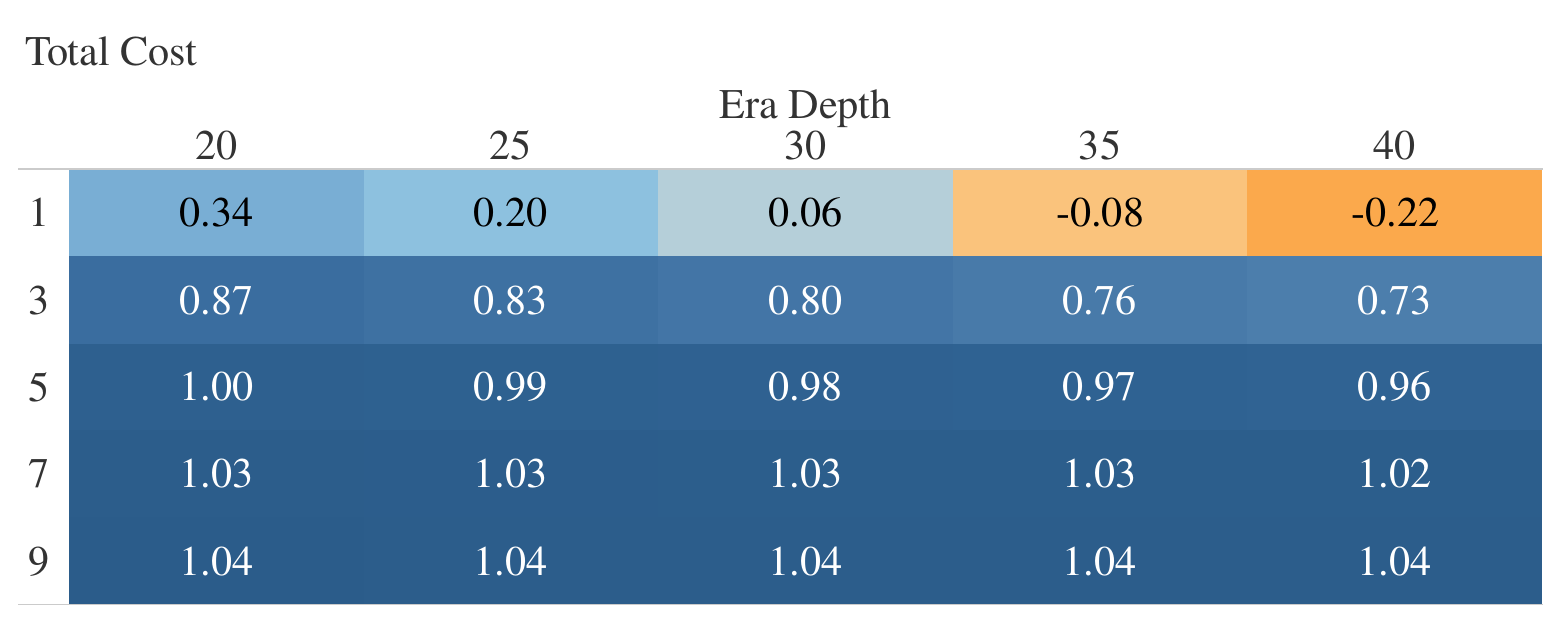}
        \caption{On-chain Gas-saving Ratio}
        \label{fig:total_cost}
    \end{subfigure}
    \caption{Figure \ref{fig:pt_circuit} illustrates the relationship between $d_{\it era}$ and $d_{\it slot}$ on the prover time for $\textit{Cir}_{\textit{dep.}}$ and $\textit{Cir}_{\textit{wit.}}$. The prover time of $d_{\it slot}=5$ and $d_{\it era}=20$ is set to be uint 1. Figure \ref{fig:total_cost} shows the variation in the reduction of the final on-chain total gas consumption as the depths of \emph{slot} and \emph{\it era} change. In the current experimental configuration, where the depths of \emph{slot} and \emph{\it era} are 5 and 20, respectively, the on-chain cost reduction percentage is taken as unit 1. A result greater than 1 indicates more savings in gas compared to the current experimental configuration, while a result less than 1 indicates higher expenses.}
    \Description{}
    \label{fig:circuits}
\end{figure}

\paragraph{Prover Time} \autoref{fig:pt_circuit} shows that the depth of the \emph{\it era} Merkle tree only affects the prover time of $\textit{Cir}_{\textit{wit.}}$. In practical deployment, considering the protocol's limitation on the number of private transactions it can support, such as a depth of 32 in \emph{TC} corresponding to $d_{\it era}$=27 and $d_{\it slot}$=5 in this protocol, more attention should be paid to the impact of the depth of the \emph{\it era} Merkle tree on prover time. Moreover, \autoref{fig:pt_circuit} indicates that the prover time of $\textit{Cir}_{\textit{wit.}}$ is less sensitive to the increase in $d_{\it slot}$ compared to $\textit{Cir}_{\textit{dep.}}$, meaning that changes in $d_{\it slot}$ mainly affect the prover time of $\textit{Cir}_{\textit{dep.}}$. Therefore, in practical production environments, the configuration of $d_{\it slot}$ needs to balance the time required for generating the $\textit{Cir}_{\textit{dep.}}$ proof and the number of private transactions batched in each batch. Shorter prover time will reduce the completion cycle of each private transaction, providing a better user experience, while more private transactions batched in each batch will reduce on-chain costs.

Under the current experimental configuration, the results of constructing Merkle trees using the same $H_{\it MiMC}$ indicate that the circuit has sacrificed more than twice the prover time compared to \emph{TC} due to more constraints. We express the total number of constraints in \emph{TC} $N_{\it TC}$ and our protocol $N_{\it New}$ in \autoref{equation4} below:\footnote{For simplicity, this formula is an approximation. The actual formula, for instance, in the case of $H_{\it MiMC}$ where the number of constraints for a single operation is 1,320, is: \( 1,815 + 1,323 \times (d_{\it slot} + d_{\it era}) + 1,320 \times (2^{d_{\it slot}} - 1) \)}
\begin{equation} \label{equation4}
N_{\it TC} = 1,815 + H_c \times depth, \quad N_{\it New}= 1,815 + H_c \times (2^{d_{\it slot}} + d_{\it era})
\end{equation}
In \autoref{equation4}, $H_c$ represents the number of constraints required for a single operation of the hash function. In the current testing environment with $d_{\it slot}=5, d_{\it era}=20$, and using the $H_{\it MiMC}$, the number of constraints for $\textit{Cir}_{\textit{dep.}}$ is 40,920, with an average prover time of 4,074ms. The number of constraints for $\textit{Cir}_{\textit{wit.}}$ is 34,890, with an average prover time of 3,956ms. These results are consistent with those in \autoref{fig:run_proof} of \autoref{bench} and meet expectations. For \emph{TC}, the number of constraints is only 34,815 with a Merkle tree depth of 25\footnote{In our current experimental configuration, the depths of \emph{\it era} and \emph{slot} are 20 and 5, respectively, meaning that up to $2^{25}$ transactions can be processed, corresponding to a Merkle tree depth of 25 in \emph{TC}.
}, with an average prover time of 3,929ms. To avoid sacrifices in prover time, we replaced $H_{\it MiMC}$ with a more efficient $H_{\it Poseidon2}$ in the circuit as the hasher, reducing the original 1,320 to 240 for a single hash operation. According to \autoref{equation4}, the number of constraints is approximately 14,295, with an average prover time of only 1,600ms, consistent with the results in \autoref{fig:run_proof} of \autoref{bench}. This optimization significantly reduces the constraints in the circuit and saves nearly 60\% of the prover time compared to \emph{TC} using $H_{\it MiMC}$.

\begin{table*}[ht]
    \centering
    \resizebox{\textwidth}{!}{
    \begin{tabular}{cccc|cccc}
        \toprule
           & \multicolumn{3}{c}{\textbf{TC}} & \multicolumn{4}{c}{\textbf{Our Protocol}} \\
          & \emph{Deposit} & \emph{Withdraw} & Total & \emph{Deposit} & \emph{Commit} & \emph{Withdraw} & Total \\
         \midrule
         \textbf{Gas} (gas) & 938,626 & 267,998 & 1,206,624 & 27,880 & 1,135,599 & 267,964 & 1,431,443\\
         \midrule
         \textbf{\emph{Ethereum}} (Gwei/ USD) & \num{9.39e6} (30.53) & \num{2.68e6} (8.71) & \cellcolor{gray!20} \num{12.07e6} (39.24) & \num{0.27e6} (0.87) & \num{0.35e6} (1.14) & \num{2.68e6} (8.71) & \cellcolor{yellow} \num{3.30e6} \\
         \textbf{\emph{Hedera}} (HBAR/ USD) & 0.96 (0.13) & 0.58 (0.08) & \cellcolor{gray!20} 1.54 (0.22) & 0.53 (0.07) & 0.03 (0.00) & 0.58 (0.08) & \cellcolor{yellow} 1.14 \\
         \textbf{\emph{BNB Chain}} (Gwei/ USD) & \num{9.39e6} (5.72) & \num{2.68e6} (1.63) & \cellcolor{gray!20} \num{12.07e6} (7.35) & \num{0.27e6} (0.16) & \num{0.35e6} (0.21) & \num{2.68e6} (1.63) & \cellcolor{yellow} \num{3.30e6} (2.01) \\
         \bottomrule
    \end{tabular}
    }
    \caption{A comparison of the gas consumption and fees for privacy transactions in our protocol VS. \emph{TC}}
    \Description{}
    \label{tab:gasfees}
\end{table*}

\paragraph{On-chain Cost} As for the on-chain cost for users, gas consumption are incurred during the deposit and withdrawal operations. Since the algorithm executed in the smart contract during withdrawal has constant space complexity, the cost of the withdrawal process remains unchanged. In the test, the gas consumption for a single withdrawal was 267,964, with approximately 200K gas consumed in the process of calling the validator smart contract. The gas consumption for a single deposit into the mixer pool by a user is 27,880.

As for the on-chain cost for the sequencer, gas consumption are only incurred during commit operations. The commit operation changes the state of the \emph{\it era} on-chain, hence exhibiting linear space complexity \(O(d_{\it era})\). In the test, the gas consumption for a single deposit was 1,157,773, with approximately 220K gas consumed in the process of calling the validator smart contract built using the \emph{Groth16} protocol. The difference in cost between commit and withdrawal operations lies in publishing the hash value of the root node of the \emph{\it era} used for batching off-chain transactions.

With the above cost test results as benchmarks, we forked the open-source code of \emph{TC} and obtained a baseline in the same testing environments. \autoref{tab:gasfees} shows the corresponding gas consumption and fees results on the mainnets of the three testnets under our protocol versus \emph{TC}.\footnote{Gas and USD in the last column are represented as amortized values, meaning that all expenses per batch are distributed over 32($= 2^5$) transactions included in each batch. At the time of writing this paper, 1 ETH = 3,251 USD, 1 BNB = 609 USD, 1 HBAR = 0.14 USD, and the gas price on the \emph{Ethereum} and \emph{BNB Chain} mainnet are both 10 Gwei.} 

In our protocol, assuming the cost of \enquote{commit} operations borne by the sequencer is evenly distributed among users (with the current test environment's $d_{\it slot} = 5$ and a corresponding transaction batch size of 32 transactions), for \emph{Ethereum} and \emph{BNB Chain}'s on-chain pricing mechanisms, the total gas consumption per transaction is approximately 331,331. This is approximately a 73\% reduction compared to \emph{TC}'s gas consumption of 1,206,624 per transaction, resulting in fees on \emph{Ethereum} and \emph{BNB Chain} being only around 27\% of the original. For \emph{Hedera}, its fee schedule is set by the \emph{Hedera Governing Council} and is always based on USD~\cite{feeEstimater}.\footnote{\emph{Hedera} provides USD estimates for the fees of all network services' transactions and queries using the API~\cite{Hederafees} and offers the \emph{Hedera} Fee Estimator~\cite{feeEstimater} for a more accurate estimation of fees for mainnet transactions and queries.}  Due to the different pricing mechanisms, we recorded the actual levels of HBAR consumed on the testnet. We found that in our protocol and \emph{TC}, the total cost per single privacy transaction was 1.14 HBAR and 1.54 HBAR, respectively, representing a reduction of nearly 26\%. \autoref{fig:total_cost} indicates that with the increase in \emph{slot} depth, the overall on-chain cost will further reduce, whereas the increase in \emph{\it era} depth will conversely increase the on-chain cost burden.



\section{Discussion}
Our work contributes to the evaluation of SNARKs implementation with ZK-friendly hash functions. However, we believe there is still potential for improvement in the breadth of benchmarking ZK tools and algorithms, as well as in the depth of exploration and evaluation of SNARKs implementation.

\subsection{A more comprehensive benchmark of ZK toolsets} With the rapid development of ZKP, not only SNARKs but also other prominent ZKP protocols like \ac{STARKs} are being introduced. Conducting a comprehensive evaluation of the corresponding ZKP development tools is a valuable challenge. We believe that future efforts can focus on the following two aspects to achieve this comprehensiveness.
\paragraph{ZKP development tools} In terms of ZKP development tools, the choice of ZK-development tools in \autoref{bench} should consider more libraries that support a wider range of ZKP protocols, such as \emph{halo2}~\cite{halo2kzg}, \emph{gnark}~\cite{gnark091}, etc. Additionally, this experimental sample should not only include SNARKs-type protocols but also cover popular STARKs-type protocols like \emph{boojum}~\cite{boojum}, \emph{starky}~\cite{Plonky2}, to provide a more comprehensive benchmark between libraries.

\paragraph{ZK-Circuit} In terms of the algorithmic implementation of ZK-circuits, as mentioned in \autoref{subsubsec: hash}, more and more ZK-friendly cryptographic algorithms are being innovated, such as type 2 and type 3 hash function algorithms. Moreover, the configuration of security constants in the implementation of these algorithms, such as the construction of round constants, should have more implementation choices beyond a single one, enabling a more reliable evaluation of the performance of these algorithms through a more comprehensive implementation. Additionally, future work should cover a wider variety of cryptographic primitives beyond hash functions, such as signature algorithms.

\subsection{A more complete SNARKs-based privacy-preserving protocol} ZKP has been continuously innovated in blockchain applications due to its inherent privacy, security, and other features. In broader application areas, beyond the popular ZKP application of privacy protection protocols chosen in this work, exploring how to unlock the potential of ZKP should be an interesting task. In the field of privacy protection protocols covered by this work, we believe that its completeness can be achieved in the following two aspects.

\paragraph{Application Scenarios} For the application scenarios of privacy protection features, broader demand groups in daily life should be considered, such as the privacy needs of healthcare data mentioned in \autoref{exploration}. In the application scenario chosen in our work, namely the ZKP mixer aimed at enabling private encrypted currency transactions, they are currently drawing attention from centralized regulators, notably the U.S. government, due to their robust privacy protection for user transactions. For this reason, we discussed in \autoref{subsec: privacy model} how to utilize the role of a sequencer to provide resilient censorship, i.e., by filtering addresses that meet certain criteria to participate through a pre-defined \emph{Non-banned list}. In future work, the establishment of these rule standards should be carefully considered to prevent addresses with suspicious transaction behavior from engaging in illegal activities such as money laundering.

\paragraph{ZKP mixer protocol} In terms of designing a more robust ZKP mixer protocol, our starting point is to reduce on-chain interaction costs, but there are many more aspects worth studying for performance extension. An ideal scenario would be that for any transaction amount, users only need one deposit and one withdrawal operation to complete the transaction, rather than splitting the transaction amount into multiple transactions based on the configuration of the mixer pool and performing multiple deposit and withdrawal operations. This improvement can greatly enhance the flexibility of user transactions and reduce costs. Additionally, more efficient ZKP protocols, such as \emph{Virgo}~\cite{virgo} and \emph{deVirgo}~\cite{zkbridge}, can be considered to further reduce the proof generation time by combining with SNARK protocols, while also increasing the protocol's throughput capacity. Furthermore, in \autoref{subsec: privacy model}, we found that enhancing the number of participated transactions per unit time can improve the privacy of the protocol. Therefore, incentives like anonymous mining can be introduced to increase the number of privacy transactions.

\section{Related Work} \label{related}

As mentioned in \autoref{subsubsec: hash}, the design concept of Circuit-Friendly Hash Functions, such as \emph{Poseidon2}~\cite{grassi2023}, \emph{Neptune}~\cite{neptune2021}, and \emph{GMiMC}~\cite{ben2020}, focuses on iterating low-degree components (referred to as \emph{rounds} in symmetric cryptography) multiple times to reach a high degree, meeting security requirements while maintaining fast computation on a CPU. Another approach reduces circuit constraints through non-procedural computation, an idea first introduced by the \emph{FRIDAY} hash function~\cite{albrecht2019algebraic}, and further explored in functions like \emph{Rescue}~\cite{aly2019} and \emph{Griffin}~\cite{grass2022}, which require fewer \emph{rounds} but result in slower plain performance. A third category of ZK-friendly hash functions, exemplified by \emph{Reinforced Concrete}~\cite{grassi2022rc} and \emph{Tip5}~\cite{alan2023}, strikes a balance by narrowing the range of values for any function, creating a \emph{lookup table} component to improve both plain and circuit performance.

Benchmarking ZKP tools for performance evaluation has seen notable efforts, including Celer Network's~\cite{daniel2020} SHA-256 circuit benchmarking for runtime and memory usage. This work was extended by Jens et al.~\cite{zkbench} with the \emph{ZK-Harness} framework, supporting a wide range of circuits, libraries, and tools. Sanjay et al.~\cite{benchmarkingcircom} introduced a \emph{Schnorr} digital signature circuit template in \emph{circom} as part of \emph{ZK-Harness}. Our work further extends this by benchmarking different proving systems and evaluating ZK-friendly hash functions' performance, providing popular circuit templates like \emph{Poseidon2}.

ZKP applications are broad, with blockchain mixers enhancing anonymity by obscuring transaction chains, inspired by privacy-preserving ZKP-based transactions such as \emph{Typhoon.Cash}~\cite{typhooncash} and \emph{Typhoon.Network}~\cite{TyphoonNetwork}. We designed a non-custodial ZKP mixer to enhance transaction privacy. Another promising application, ZK-Rollup, aggregates off-chain transactions to extend layer 1 blockchains~\cite{ArbitrumEthereum, Sguanci2021LayerSurvey}, enabling public verification of transaction correctness through data availability. Examples include \emph{zkSync}~\cite{igor2018zksync}, \emph{Starkware}~\cite{Starknet}, and \emph{Scroll}~\cite{scroll}. However, ZK-Rollup has not yet been applied to mixers. Our work pioneers the use of ZKP for off-chain transaction batching in privacy-preserving protocols, significantly reducing on-chain costs and offering elastic censorship.
\section{Conclusion} \label{conclusion}
This study benchmarks various stages of SNARKs, comparing three proving systems in snarkjs and five ZK-friendly hash functions on curve \emph{bn254}, with circuit templates for three functions in \emph{circom}. Results indicate that under the \emph{Groth16} proving system, \emph{Poseidon} and \emph{Poseidon2} are the most efficient in RAM and runtime during proving, while \emph{MiMC} has the smallest circuit size in \emph{Plonk} and \emph{Fflonk}. Using a unified test circuit based on the \emph{MiMC} hash function, \emph{Groth16} shows the best RAM and runtime efficiency during proving.

We also proposed a privacy-preserving protocol based on SNARKs, introducing a sequencer to implement batching, enhancing cost-saving and efficiency. This protocol was implemented with TC as the baseline in the context of cryptocurrency privacy-preserving transactions. This implementation demonstrated a 73\% cost reduction on EVM blockchains and 26\% on \emph{Hedera}. This improvement doubles the circuit size, but replacing \emph{Poseidon2} as the hash function for Merkle trees further reduces proving time by 60\% from the baseline. Additionally, we found that the privacy of our solution gradually increases with the number of historical transactions involved.

\section*{Acknowledgements}

This work was supported by the Ethereum Foundation through the ZK Grant Round under Grant ID: FY24-1503. We thank Arafath Shariff for his help with an early version of our \href{https://github.com/hanzeG/circom-zkmixer}{code base}.

\bibliographystyle{ACM-Reference-Format}
\bibliography{references}

\appendix
\section{Appendix} \label{appendix}

\begin{lstlisting}[language=Solidity, caption={\href{https://github.com/hanzeG/circom-zkmixer/blob/main/contracts/Protocol.sol}{Smart Contract $SC_{era}$}}, label={lst:era}]
interface IHasher {
  // in_xL, in_xR represent the pre-images
  function MiMCSponge(uint256 in_xL, uint256 in_xR) external pure returns (uint256 xL, uint256 xR);}
  
contract Era {
  constructor(
  uint32 _levels, // Era's depth
  IHasher _hasher) // Implementing MiMCSponge
  
  // Calls MiMCSponge() to perform MiMC hash
  function hashLeftRight(IHasher _hasher, bytes32 _left, bytes32 _right) public pure returns (bytes32){}
  
  // Calls hashLeftRight() to update era
  function _insert(bytes32 _leaf) internal returns (uint32 index){}}
\end{lstlisting}

\begin{lstlisting}[language=Solidity, caption={\href{https://github.com/hanzeG/circom-zkmixer/blob/main/contracts/Protocol.sol}{Abstract Smart Contract $SC_{protocol}$}}, label={lst:protocol}]
interface IVerifier {
    function verifyProof(uint[2] calldata _pA, uint[2][2] calldata _pB, uint[2] calldata _pC, // Relayer/user ZKP
    uint[1] calldata _pubSignals // Public signals of deposit and withdraw circuit
    ) external returns (bool);}

abstract contract Protocol is Era, ReentrancyGuard {
  constructor(
    IVerifier _verifier1, // Relayer ZKP verifier
    IVerifier _verifier2, // User ZKP verifier
    IHasher _hasher, // Implementing MiMCSponge
    uint256 _denom, // Denomination
    uint32 _eraDepth // Era's depth
  ) Era(_eraDepth, _hasher)

  // Record user's commitment and accept deposits
  function deposit(
    bytes32 _userCommitment // User commitment
  ) external payable nonReentrant {}

  // Call _insert() to update era
  function commit(uint[2] calldata _pA, uint[2][2] calldata _pB, uint[2] calldata _pC, // Relayer ZKP
  uint[1] calldata _pubSignals, // Public signals of deposit circuit
  bytes32 _relayerCommitment // Root of slot
  ) external payable nonReentrant {}

  // Verify user's identity and security
  function withdraw(uint[2] calldata _pA, uint[2][2] calldata _pB, uint[2] calldata _pC, // User ZKP
  uint[1] calldata _pubSignals, // Public signals of withdraw circuit
  bytes32 _root, bytes32 _nullifierHash, address payable _recipient, address payable _relayer, uint256 _fee, uint256 _refund // Elements in the beta array used for authentication and security checks.
  ) external payable nonReentrant {}
}
\end{lstlisting}

\section{Ethics}

There is a risk of our improvements to the ZKP mixer protocol being exploited for illicit purposes. We strongly declare our firm opposition to the illicit use of ZKP mixers for activities such as money laundering. We outline preventive measures regarding this issue in \autoref{subsec: privacy model}.

\end{document}